\documentclass[aps,prf, twocolumn]{revtex4-1}
\usepackage{amsmath,amsfonts,bm, color,amssymb}
\usepackage{graphicx,epsfig,overpic,hyperref,ulem}%% OPTIONAL MACRO DEFINITIONS
\input epsf
\UseRawInputEncoding

\usepackage{graphicx}
\usepackage{epstopdf}
\usepackage{graphics}
\usepackage[euler]{textgreek}

\newcommand{\refeq}[1]{Eq. \ref{#1}}

\newcommand{\Cm}{\bar C}

\newcommand{\kT}{ k_BT}

\newcommand{\sigm}{{\sigma}}
\newcommand{\eps}{{\varepsilon}}
\newcommand{\out}{{\mathrm{w}}}
\newcommand{\ins}{{\mathrm{v}}}
\newcommand{\mem}{{\mathrm{m}}}

\newcommand{\ehd}{{\mathrm{d}}}
\newcommand{\el}{{\mathrm{el}}}

\newcommand{\Rr}{{\it \Lambda}}
\newcommand{\Sr}{{\it S}}

\newcommand{\eq}{{\mathrm{eq}}}
\newcommand{\visrat}{{\chi}}

\newcommand{\im}{{\mathrm i}}

\begin{document}

\title{A vesicle microrheometer for high-throughput viscosity measurements  of lipid and polymer membranes}
\author{Hammad A. Faizi$^1$, Rumiana Dimova$^2$ and Petia M. Vlahovska$^3$}

\affiliation{$^1$ Department of Mechanical Engineering, Northwestern University, Evanston, IL 60208, USA\\
$^2$Department of Theory and Biosystems, Max Planck Institute of Colloids and Interfaces, Science Park Golm, 14424 Potsdam, Germany\\
$^3$Department  of Engineering Sciences and Applied Mathematics, Northwestern University, 60208, USA, email: petia.vlahovska@northwestern.edu}


\begin{abstract}

Viscosity is a key  property of cell membranes that controls mobility of embedded proteins and membrane remodeling. Measuring it is challenging because  existing approaches involve complex experimental designs and/or models, and the applicability of some is limited to specific systems and membrane compositions. As a result there is scarcity of data  and the reported values  for membrane viscosity vary by orders of magnitude for the same system. Here, we show how viscosity of  bilayer membranes can be obtained   from the transient deformation of giant unilamellar vesicles.  The approach enables a non-invasive,  probe-independent and high-throughput measurement of the viscosity of  bilayers made of  lipids or polymers with a wide range of compositions and phase state. Pure lipid  and single-phase mixed bilayers are found to behave as Newtonian fluids with strain-rate independent viscosity, while  phase-separated and diblock-copolymers systems exhibit shear-thinning  in the explored range of strain rates 1-2000 $s^{-1}$.  The results also reveal that electrically  polarized bilayers can be significantly more viscous than charge-neutral bilayers.    These findings suggest that biomembrane viscosity is a dynamic property   that can be actively modulated  not only by composition but also by membrane polarization, e.g., as in action potentials.

\end{abstract}

\date{ \today}

\maketitle

Cells and cellular organelles are enveloped by membranes,
whose main structural component is a lipid bilayer  \cite{Singer1972}.  The lipid bilayer 
%is made from non-covalently bound lipid molecule that can rearrange and thus
endows membranes with fluidity that is essential for 
functions that depend on biomolecules mobility, e.g., signaling
\cite{Cohen:2019,LeRoux:2019,Illukkumbura:2020}. Fluidity is 
%assumed to be 
modulated by membrane composition and this homeoviscous adaptation is crucial for the survival of organisms that can not regulate their body temperature like bacteria  \cite{Sinensky,Hazel:1995}. 
%Despite its importance, however, quantification of membrane fluidity remained elusive. 
Viscosity is the common measure for fluidity, yet for membranes this property has been challenging  to assess.  Data for viscosity of lipid membranes are limited and reported values vary significantly, sometimes by orders of magnitude for the same system (SI Table 1).  
For example,  reported values for the surface shear viscosity  of membranes made of a typical lipid such as dioleoylphosphatidylcholine (DOPC) span two orders of magnitude: $(0.197 \pm 0.0069)\times10^{-9}$ Pa.s.m \cite{Zgosrki:2019}, $(1.9\pm11) \times10^{-9}$  Pa.s.m \cite{Honerkamp:2013}, $(16.72\pm1.09)\times 10^{-9}$ Pa.s.m  \cite{Chakraborty21896}. For a similarly structured lipid, palmitoylphosphatidylcholine  (POPC), the surface viscosity measured by shear rheology of  Langmuir monolayers  is $3\times 10^{-4}$ Pa.s.m \cite{Espinosa6008}.
%Shear rheology of Langmuir monolayers \cite{Espinosa6008} obtains much higher viscosities 
%%unusually high value for viscosity of the typical membrane lipid POPC
% than the values measured by methods that use bilayer membranes, a subject of ongoing debate \cite{Guzman:2018}.
 %  {The composition of monolayer would depend on the affinity of the molecule to the interface.}
%which questions the equivalence of lipid monolayer and bilayers Langmuir monolayer are not representative
%planar bilayers spanning apertures in supports, also known as black lipid membranes (BLMs)
%?? is BLM free-standing membrane???
Experimental methods  that utilize free-standing bilayer membranes, e.g, vesicles or black lipid membranes,  rely on estimates from  the rate of tether formation \cite{WAUGH198229}, diffusion coefficients of domains  \cite{Cicuta:2007,Petrov:2012,Block:2018} or membrane-anchored nanoparticles \cite{Herold:2013,Hormel:2014},  domain shape fluctuations \cite{Camley:2010}, domains motion on vesicles induced by applied flow \cite{Honerkamp:2013,Imai:2020}, bilayer thickness fluctuations or lipid dynamics 
%of lipid acyl tails 
%at the structural correlation peak
measured with neutron spin echo spectroscopy \cite{Nagao:2017, Kelley:2020,Nagao:2021}, 
 %fluorescence lifetime imaging of small synthetic molecules called  ``molecular rotors" embedded in the membrane \cite{Kuimova:2013}, 
% absorbance and fluorescence emission  excited-state li
 fluorescence quantum yield or  lifetime of viscosity-sensitive fluorescent dyes \cite{Kuimova:2013,Chwastek}, and the forced motion of colloidal particles in the membrane \cite{Velikov:1997,Dimova2000, Dimova1999}.  In silico approaches, using molecular dynamics simulations, have also been developed to determine
membrane viscosity  \cite{Shkulipa:2007,Vogele:2018,Zgosrki:2019}. 
Despite these advances,  the systematic study of membrane viscosity has been hindered by various limitations of the proposed methodologies. 
For example, domain-based methods \cite{Cicuta:2007,Petrov:2012,Camley:2010,Honerkamp:2013,Imai:2020} are limited to phase-separated membranes  and the measured  viscosity reflects the continuous phase, not the membrane as a two-phase fluid. Bilayer thickness fluctuations \cite{Nagao:2017}  depend on both shear and dilational monolayer viscosities. In particle-based methods \cite{Herold:2013,Hormel:2014,Velikov:1997,Dimova2000, Dimova1999}, the probe perturbs the  membrane and the data interpretation requires complicated analysis that discerns 
the contributions to the particle mobility from the flow in the membrane and the surrounding fluids \cite{Gurtovenko:2019, Danov-Dimova:2000,Naji:2007}. Furthermore, since membrane surface viscosity is a macroscopic quantity, defined on scales where the bilayer can be modeled as a two-dimensional incompressible fluid,  methods utilizing measurements  at the micro- or  nano- scale and/or based on molecular probes   may not report the effective continuum viscosity  but a quantity, often called "microviscosity", which is local and depends on the immediate environment  \cite{Nagao:2021}.
These complexities are likely  the source of the huge variability in reported values of viscosity for lipid bilayer membranes, making it  challenging to  compare data obtained by different methods. 

\begin{figure*}[t!]
\centering
\includegraphics[width=\linewidth]{{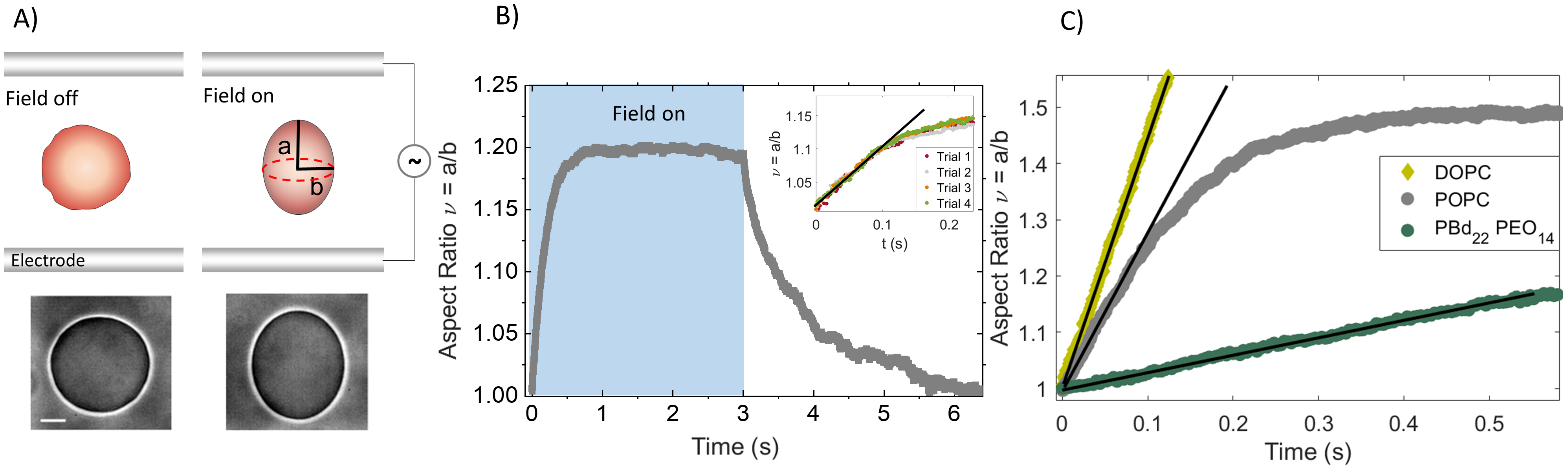}}
\caption{  \footnotesize{ Electrodeformation method to measure membrane viscosity. (A) A uniform electric field  deforms a GUV into a prolate ellipsoid by pulling out area stored in suboptical thermally-excited membrane undulations.  Snapshots of the vesicle during the experiment.
Imaging with phase contrast microscopy. Scale bar: 15 $\mu$m.
(B)  Prolate deformation of a POPC GUV in an electric field with amplitude $E_0$= 10 kV/m and frequency
of 1 kHz. Time zero in all graphs corresponds to turning the field on.The inset shows that repeated deformation does not alter the initial slope of the deformation curve.
C) Vesicles made of lipids (DOPC, POPC) and the diblock-copolymer PBD$_{22}$PEO$_{14}$ deform at a different rate indicating different membrane viscosity. The field strength and frequency are 8 kV/m and 1 kHz. 
 The solid lines correspond to the theoretical fit  with   \refeq{EqV}.  
   }}
\label{fig1}
\end{figure*}

 Here, 
we show that the deformation of giant unilamellar vesicles (GUVs) can be employed to obtain the surface "macroviscosity", i.e., the shear viscosity of the membrane treated as a two-dimensional incompressible fluid.
%of both homogenous and phase-separated membranes. 
% previously used to measure bending rigidity of membranes \cite{kummrow1991,Gracia:2010,yu2015}, 
 Upon application of an extensional stress, e.g., generated by an uniform electric field  \cite{Riske-Dimova:2005,Aranda:2008,Salipante-Vlahovska:2014,Lin-Dimova:2015}, extensional flow \cite{Deschamps-Steinberg:2009,Shenoy:2016,Dinesh:2020}, or an optical stretcher \cite{Guck:2001,Guck:2015}
%a uniform AC electric field, 
a quasi-spherical vesicle deforms into a prolate ellipsoid.
The aspect ratio, $\nu=a/b$ (see sketch in  Fig. \ref{fig1}A), increases and reaches a steady state. When the stress is removed, the vesicle relaxes back to its equilibrium spherical shape, see Fig. \ref{fig1}B. The reproducibility of the results was tested by repeated measurements with the same vesicle, see Fig. \ref{fig1}B inset, showing identical slope of the aspect ratio curves for small deformations.

Even though the applied stress is extensional and vesicle deformation is axisymmetric,  material transported on the vesicle surface  undergoes shear because the membrane is area--incompressible \cite{Henle:2009, woodhouse_goldstein_2012,Rahimi:2013, Sigurdsson:2016,Vlahovska:2019book}.
%%The rate at which the vesicle deforms and its aspect ratio $\nu=a/b$ (see sketch in Fig. \ref{fig1}A) changes %provides a direct measure of
%depends on  the membrane shear viscosity. 
%Note that the 
%increase in apparent area comes from ironing of suboptical membrane undulations, and thus dilatational 
The rate at which the vesicle elongates while the field is on, and relaxes back to its equilibrium shape after  the field is turned off, is related to the membrane shear viscosity.
For small deformations,  $\nu\lesssim1.3$, 
the evolution of the aspect ratio is described by  \cite{Vlahovska:APBL,Salipante-Vlahovska:2014,Vlahovska:Stone,Vlahovska:2019book} (see Appendix Section 2 for a summary of the theory)
%\cite{Vlahovska:APBL,Vlahovska:Stone,Vlahovska:2019book}
\begin{equation}
%\nu(t)=1+\frac{p R}{\sigma(\nu)}\left(1-\exp{\left(-\frac{24\sigma(\nu)}{\eta R\left(55+16\visrat_m\right)}{t}\right)}\right)
\dot\nu=\frac{1}{\eta\left(55+16\visrat_m\right)}\left(p -\frac{24 \sigma}{R} (\nu-1)\right)
\label{shapeev}
\end{equation}
where $\visrat_m=\eta_m/\eta R$ is the dimensionless  surface viscosity $\eta_m$, $\eta$ is the viscosity the solution inside and outside the vesicle (assumed to be the same), $\sigma$ is the membrane tension,
%, which depends on the instantaneous deformation, 
and $R$ is the vesicle radius. 
%Bending stresses are negligible to tension, $\kappa/\sigma R^2\ll 1$.
 In an applied extensional flow with strain rate $\dot\gamma$,  $p=180\eta \dot\gamma$. In the case of a charge-neutral vesicle in a uniform DC electric field with amplitude $E_0$, $p=27\eps E_0^2/4 $, and in an AC field with frequency $\omega$, $p(\omega)$ is given in Appendix Section 2.
 %  (note that even in an AC field the electric stress has a steady component because the stress is quadratic in the electric field). 
 Thus from the vesicle dynamics in response to an  extensional  flow or a uniform electric field it is straight-forward to obtain the membrane viscosity.
 Note that the dynamics does not involve dilational viscosity because vesicle deformation and the accompanying increase in apparent area comes from ironing of suboptical thermally-excited membrane undulations, while the area per lipid remains the same.
 
% is quadratic in the electric field  and thus even an AC field it has a steady component responsible for the vesicle ellipsoidal deformation).
% The theoretical model and the limitations of the approximation \refeq{EqV} are discussed in the Supplementary Material.
%The theoretical model is summarized  in theAppendixSection 2.

\section*{Results}

%  The use of the AC field  is preferable to DC, because DC pulses often cause membrane  poration and  the vesicle can form sharp edges \cite{Riske-Dimova:2005,Riske-Dimova:2006}, which complicate the shape analysis.  

We illustrate the implementation of the approach on the example of a quasi-spherical vesicle subjected to a uniform  AC  electric field.  The electric field is advantageous over extensional flow or radiation pressure in optical stretchers because of  (i) the simplicity of the experimental set up, and (ii) the ability to create  potential  difference across the bilayer emulating transmembrane potentials in living cells. The use of an AC field is preferable because (i)  a DC field
%, even on the order of few ms, 
could cause  Joule heating and  electroosmotic flows, whose
% are also common in  DC electric fields and their
effect on vesicle deformation and stability is difficult to account for  \cite{Lacoste:2009}, and (ii) the transmembrane potential can be modulated by changing the electric field frequency \cite{Grosse-Schwan:1992,Vlahovska-Dimova:2009}.   We applied the method to fluid membranes composed of the phosphatidylcholine lipids palmitoyloleoyl (POPC), dioleoyl (DOPC), oleoylmyristoyl (OMPC),  stearoyloleoyl (SOPC), and dipalmitoyl (DPPC), cholesterol (chol) and diblock copolymers, PBd$_{x}$-$b$-PEO$_{y}$ with varying hydrophobic molecular weight, M$_h$, from 0.7 kDa to 6.8 kDa.  

% and  a component that oscillates with twice the applied frequency. [ADD THE plot with the oscillations in the SI]  COULD there be  a viscouelatic effet plot the electric shear as a function of frequcny, increase and decrease, no plateaus? so that excludes ]
%Furthermore, AC field suppresses vesicle electrophoresis making the method applicable to bilayers containing charged lipids.

  \begin{figure*}[t!]
\centering
\includegraphics[width=\linewidth]{{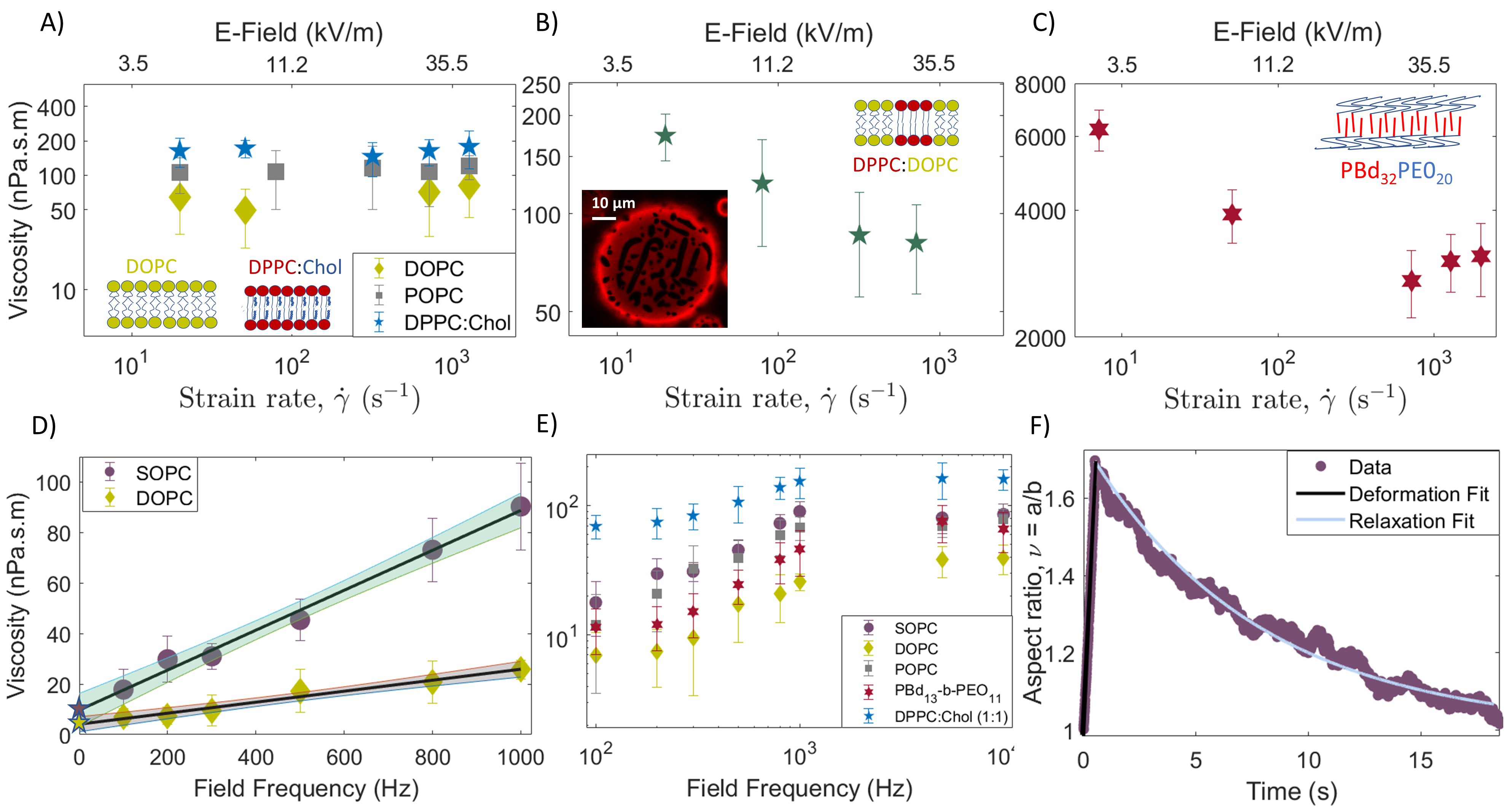}}
\caption{  \footnotesize{Viscosity dependence on electric field strength, or equivalently strain rate $\dot\gamma=\eps E_0^2/\eta$,   and  frequency.  A) Single-component (DOPC or POPC) and  single-phase multicomponent bilayers (DPPC:Chol 1:1) behave as Newtonian fluids. B) Viscosity of the phase-separated multicomponent (DPPC:DOPC 1:1) decreases with the effective strain rate. The inset shows an intricate network of finger-like domains indicative of fluid (red) and gel (dark) phase coexistence. C) Diblock-copolymer  bilayers also display shear thinning viscosity.  All  measurements  in panels A-C, were done at field frequency 1kHz.
D) Viscosity measured at different frequencies at a fixed strain rate $\dot\gamma=50\,s^{-1}$, corresponding to field strength 8 kV/m, increases with frequency. Extrapolating to zero frequency (the DC limit) yields the shear surface viscosity. The symbols on the vertical-axis-intercept refer to the extrapolated values of 4.11$\pm2.63$ nPa.s.m and 9.73$\pm5.80$ nPa.s.m for DOPC and SOPC, respectively. The R-squared value of the linear fit is 0.98 and 0.97 for the SOPC and DOPC, respectively.  The shaded band represents a 95\% confidence interval. E) The viscosity reaches as plateau at frequencies above few kHz. This behavior is common to all studied compositions. F) Shape relaxation after the field is turned off yields the similar value for the viscosity as the zero-frequency limit obtained from the frequency sweep.   Data for SOPC,  $E_0=8$ kV/m and $\omega=1$ kHz. Tension obtained from flickering spectroscopy is 5.7$\pm 2.6\times10^{-9}$ N/m. Viscosity obtained from initial deformation and relaxation is 164 nPa.s.m and 21.1$\pm$ 29.4 nPa.s.m, respectively.}}
\label{fig2}
\end{figure*}

\subsection*{Vesicle transient deformation yields membrane viscosity.}

Figure~\ref{fig1} summarizes the experiment.  
%Upon application of the electric field, the vesicle aspect ratio increases and reaches steady state. 
%The steady shape is determined by the balance of membrane tension and applied stress. 
%We prepared  giant unilamellar vesicles (GUVs)  with various membrane compositions. 
 The elongation curves of a GUV  initially show linear increase (Figure~\ref{fig1}C and Appendix Figure S3). 
 %s of the same GUV at different field strengths initially show linear increase, see Appendix Figure S2. 
 %This slope is independent of membrane tension and only depends on membrane viscosity.
The linear slope is predicted by \refeq{shapeev} if the 
 second term, which describes the action of the tension opposing the deformation,  is neglected
%, which corresponds to shape relaxation driven by membrane tension
% shows that initially the aspect ratio increases linearly with time
 \begin{equation}
 \label{EqV}
 \nu
% =1+\frac{t}{t_{\ehd}}\left(\frac{27}{8\left(55+16\visrat_m\right)}\right)
=1+\frac{t}{t_{\ehd}}\left(\frac{p\omega)}{\left(55+16\visrat_m\right)}\right)
 \end{equation}
 where $1/t_{\ehd}=\eps E_0^2/\eta$ is 
 %the characteristic scale for the  rate of deformation for 
 the characteristic rate-of-strain imposed by the electric field.
 % ($1/t_{\ehd}=\dot\gamma$ for the extensional flow).  
 Thus, the slope of the aspect ratio plotted as a function of the rescaled time $t/t_\ehd$, at the same field frequency, depends solely on the membrane viscosity.  Vesicles made of different lipids or diblock-copolymers show different initial slopes, see Fig. \ref{fig1}C, indicating different membrane viscosities. 
 %the membrane viscosity can be obtained from the slope of the $\nu(t)$ curve immediately after the application of the field. 
 The linear response is observable only if the restoring force of the membrane tension
  is negligible compared to the deforming electric stress, i.e., $\sigma/R \eps E_0^2\ll1$, which is indeed  the case for typical values of the applied electric field $E_0=5$ kV/m, equilibrium membrane tension $\sigma=10^{-8}$ N/m, and vesicle radius $R=10 \,\mu$m. 
% However, if the tension is high, the relaxation  becomes operational as soon as the field is applied and  one needs to fit the entire deformation curve including the plateau; this situation arises when applying the method to drops  (SI Section 5).
 %REMOVE THE DROP STUFF. Hammad: I would say that this approach shows the versatility of the method to translate to different systems and droplets help achieve that. PNAS like broader stuff and this would be appreciated. Plus previous reviewers liked it too.}}. 
 The  time up to which the linear approximation is reasonable (see Appendix Section 2B) is estimated to be ${t_c}/{t_\ehd}\sim{\eps E_0^2 R\left(55+16\visrat_m\right)}/{(12\sigma)}$. 
%Considering typical parameters  and a  more conservative time estimate using $0.1 {t_c}/{t_\ehd}$, yields  $t/t_\ehd\sim 5$ if $\visrat_m\sim 0$. Higher membrane viscosity extends the linear deformation regime, if $\visrat_m\sim 10$, as seen in polymersomes made of PBd$_{22}$-$b$-PEO$_{14}$, where the cut-off time becomes $t/t_\ehd\sim15 $, see Fig. \ref{fig1}C.  
Using times up to $0.1 {t_c}/{t_\ehd}$ (or equivalently $\nu$ up to 1.1) minimizes the error in the linear fit.

\subsection*{Viscosity shows dependence on field amplitude and frequency.}
% reveals non-Newtonian rheology and charge thickening.}
%\paragraph{Bilayers as two-dimensional non-Newtonian fluids.}

The  stress, generated by the electric field,  shears the membrane with a characteristic rate $\dot{\gamma}=1/t_{\ehd}=\eps E_0^2/\eta$.
Modulating  the field amplitude thus enables us to vary the shear rate in a wide range and to probe if bilayers behave as Newtonian fluids: with  shear-rate independent viscosity.
 Increasing $E_0$ from 1 to 50 kV/m at a given frequency increases the effective shear rate  from 1 $s^{-1}$  to  2000 $s^{-1}$. We find that bilayers made of only one lipid or a homogeneous mixture, either in the liquid disordered (e.g., DOPC or POPC) or liquid ordered (e.g., DPPC:Chol 1:1) state, exhibit rate-independent viscosity (Fig. \ref{fig2}A). 
 Phase-separated bilayers such as DPPC:DOPC (1:1), which have solid domains coexisting with liquid-disordered continuous phase (see inset of \ref{fig2}B) \cite{Keller:2002,Veatch-Keller:2005,Uppamoochikkal2010},   and the diblock-copolymer membranes shear thin, i.e., their viscosity decreases with increasing shear rate (Fig. \ref{fig2}B-C and Appendix Section 3).

Intriguingly, measurements at the same field amplitude but different frequencies revealed  an apparent increase of viscosity with frequency (Fig. \ref{fig2}D). More precisely, the viscosity shows an initial linear increase with frequency and a plateau above few kHz (Fig. \ref{fig2}E).  The behavior is observed with all studied compositions. %suggesting universal, composition-independent .
%Intriguingly, measuring the viscosity at the same field amplitude but different frequencies revealed  an apparent increase of viscosity with frequency (Fig. \ref{fig2}D). More precisely,  the viscosity measure at frequencies above few kHz is almost ten times larger that the zero-frequency limit obtained fro
%while the viscosity measured at frequencies above few kHz  is constant, it decreases as the frequencies is decreased 
%We hypothesize that the membrane viscosity corresponds to the zero-frequency one
Viscosity values in the literature obtained by methods that do not involve electric fields (Appendix  Table II) are at least a factor of 10 smaller than the plateau-value of the viscosity but comparable to the zero-frequency limit obtained from the linear extrapolation at low frequencies. We hypothesize that the zero-frequency viscosity is representative for the viscosity of the membrane in the absence of electric field.  
%To test if the viscosity increase is due to the electric field, 
As a test, we analyzed the vesicle relaxation back to its equilibrium shape, after the field is turned off (Fig. \ref{fig2}F). That analysis, however, is complicated by the fact that unlike the initial elongation upon application of the field, the relaxation  is driven by and thus depends on the membrane tension. In  order to  leave the membrane viscosity as the only fitting parameter for the relaxation curve, the tension needs to be independently determined. In this experiment, the tension was obtained from the analysis of the equilibrium thermally-excited  membrane undulations (flickering spectroscopy) in the absence of electric field
% (analysis of the equilibrium thermally-driven membrane undualations) 
 before the electrodeformation experiment. The fit of the relaxation curve with \refeq{shapeev}  yielded viscosity  21.1$\pm$ 29.4 nPa.s.m. The large error is due to the uncertainty in the tension.  The viscosity value is close to the zero-frequency limit of the viscosity obtained from the frequency sweep ( 9.73$\pm5.80$  nPa.s.m) for SOPC.  We adopt the frequency sweep to determine the initial linear dependence of the viscosity and its ``no-field" limit. 
 %This has the advantage over the analysis of the shape relaxation because it is faster to carry out as it does not involve the laborious flickering spectroscopy. It is also more accurate considering the large error in the tension. That enables the analysis of 10-50 vesicles per viscosity value.
 %The frequency sweep is faster to carry out, compared to the relaxation + flickering 

To summarize, the method involves measuring apparent viscosities at different frequencies in the range 0.1-1 kHz and extrapolating to zero-frequency (as in Fig.\ref{fig2}D) to obtain the value of the viscosity in the absence of electric field.  Electric field of 8 kV/m (strain rate 50 $s^{-1}$) produces  a good range of data in the initial linear deformation regime. Measurement and data analysis of one vesicle typically takes about 10 min. We analyze 10-50 vesicles per viscosity value.
The viscosity obtained for 14 different bilayer compositions is reported in Table \ref{diffcondmethod2}.
% All measurements are done at the same strain rate, corresponding to $E_0$=8kV/m.
%effect in not related to the field amplitude but to its frequency
The values are in the range reported in previous studies (Appendix Table II). For example, the bilayer viscosity for SOPC is in good agreement with previously reported values from 3 to 13  nPa.s.m \cite{Dimova1999,Gambin2098,Shkulipa:2007}.

\begin{table}[h]
\small
  \caption{Membrane viscosity and bending rigidity  for various bilayer systems at  25.0 $^o$C and $E_0=8$kV/m (strain rate $\dot\gamma=50$s$^{-1}$).}
  %Measurements done at $E_0=8$kV/m  and temperature 25.0 $^o$C. The error in viscosity  is determined by computing the standard deviation over the vesicle population, typically 20.}
  \label{diffcondmethod2}
  \begin{tabular}{lll}
      \hline
  Composition& Viscosity & Bending Rigidity   \\
&  (nPa.s.m)&($\kT$)\\
    \hline
    DOPC& $4.11\pm2.63$&22.2$\pm2.0$\\
  OMPC& $7.73\pm3.09$&27.1$\pm2.6$ \\
  POPC & $9.32\pm5.95$&27.8$\pm2.3$ \\
   SOPC & $9.73\pm5.80$&30.1$\pm3.1$\\
  DOPC:Chol & $7.00\pm4.77$&27.8$\pm4.6$ \\
DPPC:DOPC:Chol (1:1:1) & $17.7\pm3.06$&72.0$\pm8.4$ \\
 DPPC:DOPC:Chol (1:1:2) & $15.4\pm2.40$&69.2$\pm7.9$ \\
 DPPC:Chol (1:1)  & 56.4$\pm4.63$&121.3$\pm11.0$ \\
  PBd$_{13}$-$b$-PEO$_{11}$ & $14.4\pm4.40$&17.1$\pm1.5$ \\
  PBd$_{22}$-$b$-PEO$_{14}$ & $686\pm51.0$ &31.0$\pm5.1$\\
   PBd$_{33}$-$b$-PEO$_{20}$ & $2890\pm670$ &54.4$\pm6.4$\\
  PBd$_{46}$-$b$-PEO$_{24}$ & $20600\pm4700$&NA\\
PBd$_{54}$-$b$-PEO$_{29}$ & $46700\pm900$&154.0$\pm16.0$ \\
 PBd$_{120}$-$b$-PEO$_{78}$ & $157000\pm54500$&NA \\
    \hline
    \centering
  \end{tabular}
\end{table}

\subsection*{Viscosity correlation with  lipid diffusivity, membrane composition and  thickness.}

%imaging with confocal microscopy showed that  pure DOPC, 1:1 DOPC:Chol, 1:1 DPPC:Chol, 1:1:1 DOPC:DPPC:Chol, 1:1:2 DOPC:DPPC:Chol exhibited homogeneous fluorescence. 1:1 DOPC:DPPC demonstrated phase separation with intricate network of finger-like domains indicative of gel or solid phase, while  
%  with liquid disordered state at equilibrium. 
 %2:1:1 DOPC:DPPC:Chol, formed 1-2 $\mu$m circular domains. 

% Table \ref{system} lists the membrane viscosities obtained with our electrodeformation method. Pure DOPC exhibits the lowest viscosity. 
\begin{figure}[h]
\includegraphics[width= \columnwidth]{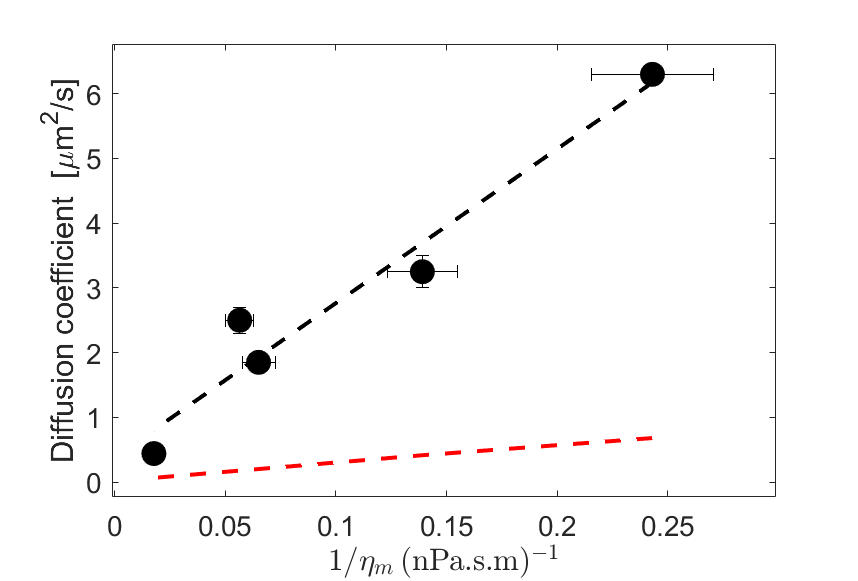}
\centering
\caption{ Membrane viscosity  as a function of diffusivity values obtained with FCS  \textit{Scherfeld et al.} \cite{SCHERFELD20033758} for membrane compositions DOPC, DOPC:Chol (1:1), DPPC:DOPC:Chol (1:1:2), DPPC:DOPC:Chol (1:1:1) and DPPC:Chol (1:1). Values of the diffusion coefficient are listed in the SI. The red dashed line corresponds to the prediction from the Saffman-Delbr{\"u}ck's model with probe radius $r=0.5$ nm. The black dashed line is a linear fit with intercept 0.383 $\mu^2$m/s and slope  23.83 $\mu$m$^3$.mPa }
\centering
\label{figD}
\end{figure}

\begin{figure*}[t]
 \includegraphics[width=\linewidth]{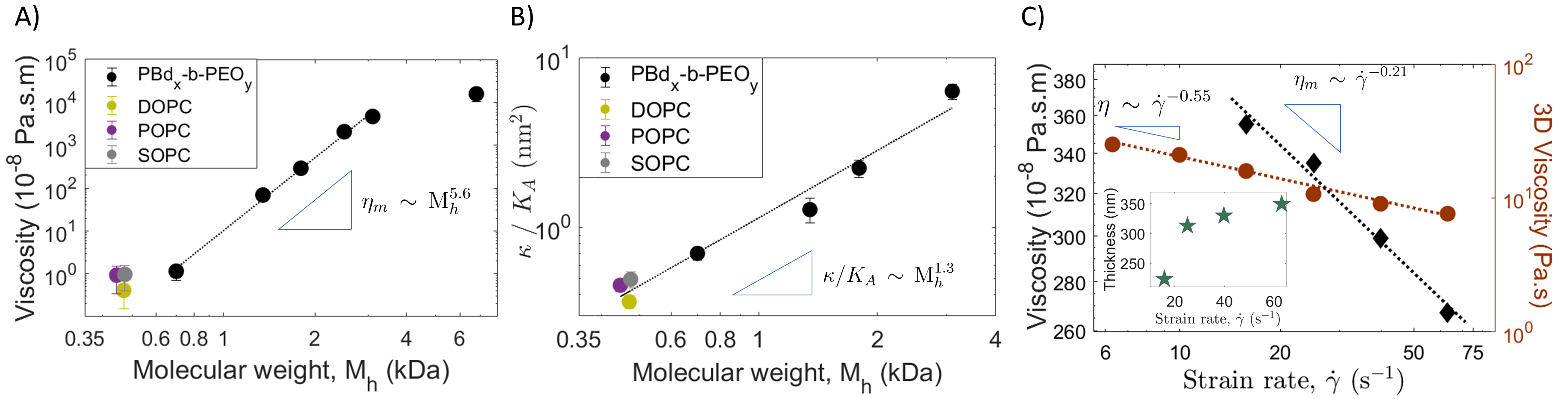}
 \centering
 \caption{A) Viscosity of PBd$_{x}$-$b$-PEO$_{y}$ bilayers as a function of the $M_h$ of the hydrophobic part (PBd). Viscosities of phospholipids are also shown for comparison. B) Bending rigidity  of PBd$_{x}$-$b$-PEO$_{y}$ and phospholipids bilayer membranes as a function of molecular weight. The power law obtained is $\kappa/K_A\sim M_h^{1.3}$. C) Comparison of the shear-thinning of the 2D and 3D viscosities of PBd$_{33}$-$b$-PEO$_{20}$. The inset shows the membrane thickness obtained from  $\eta_m/\eta=h$.}
 \centering
\label{polymer}
 \end{figure*}

Since mobility of lipids or domains is often used to asses membrane fluidity \cite{Cicuta:2007,Petrov:2012,Block:2018}, we have compared our results for membrane viscosity with measurements of   the diffusion coefficient of a lipid dye (DiI-C18) using Fluorescence Correlation Spectroscopy (FCS)  in the same ternary system \cite{SCHERFELD20033758}. The diffusion coefficient scales inversely with surface viscosity (Fig. \ref{figD}), a trend expected from the Saffman-Delbr{\"u}ck's model \cite{Saffman1975}, $D=\kT/(4 \pi \eta_m)\left(\log(\eta_m/(\eta r))-0.5772\right)$. However, there is a quantitative disagreement: using lipid dye radius  $r=0.5$ nm  (comparable to the radius of DOPC estimated from the area  per lipid head, 67.4-75.4 $\AA^2$ \cite{TRISTRAMNAGLE1998917,NAGLE2000159,Liu2004,Kucerka2006,Kucerka2008,pan2008}) the Saffman-Delbr{\"u}ck's equation predicts much lower diffusivities. 
These results suggest that while increasing viscosity does correlate with decreasing diffusivity, it is  not trivial to relate membrane viscosity and  the diffusion constant {because diffusion of molecular probes is sensitive to the probe itself as well as the bilayer structure \cite{WAUGH198229, Block:2018, Nagao:2021}. However, the empirically obtained dependence in Fig. \ref{figD}, could presumably be used to roughly deduce membrane viscosity from molecular diffusivity measurements.

The viscosity values complied in Table \ref{diffcondmethod2}  highlight  that in pure lipid systems membrane viscosity decreases with the number of unsaturated bonds in the hydrophobic tail. POPC, OMPC and SOPC have a single unsaturated bond, while DOPC has two double bonds in the hydrophobic tails. DOPC bilayers exhibit much smaller viscosity  than the other single unsaturated lipid bilayers.
% DMPC exhibits the highest viscosity, followed by POPC, SOPC, OMPC and DOPC in order .  DMPC, POPC and DOPC have none, one and two unsaturated bond in their hydrophobic tails, respectively. 
In the mixed systems, adding DPPC or/and cholesterol to DOPC  increases the viscosity. Adding cholesterol (molar ratio 1:1) to DOPC increases the bilayer viscosity as cholesterol increases the packing of the liquid disordered ($L_d$) phase. For liquid ordered ($L_o$) phase, such as DPPC:Chol (1:1), the bilayer viscosity is much higher due to the tight packing provided by saturated acyl chains. The effect is less pronounced in the ternary $L_o$ system  DOPC:DPPC:Chol (1:1:2) and DOPC:DPPC:Chol (1:1:2). For a phase-separated system of coexisting solid and fluid phases DOPC:DPPC (1:1), see Fig. \ref{fig2}B, the viscosity also increases relative to pure DOPC. The increase 
%The modest increase in viscosity  for the DOPC:DPPC phase separated mixture, see Fig. \ref{fig2}B 
agrees with an estimate of the viscosity of a 2D suspension, $\eta_{eff}=\eta_{DOPC}(1+2\phi)$, where $\phi\sim 0.4$ is the fraction of the solid phase \cite{Camley:2019}.

Using our method we were also able to  examine the commonly assumed relation between the 2D ($\eta_{m})$ and 3D viscosities ($\eta_{3D}$), $\eta_{m}= \eta_{3D}h$, where $h$  is the membrane thickness. The thickness  of  diblock-polymer bilayers
%can be varied in a wide range for block-copolymers. In these bilayers 
 %thereby offering an opportunity to test the viscosities relation. 
%The thickness of  diblock polymers bilayers
 increases with the molecular weight  of the hydrophobic part, $h\sim M_h^n$ \cite{Discher:2006,LoPresti}, where  $n$  lies within the theoretical bounds of 0.5 (random Gaussian coil) and 1 (full stretch). 
 The polymers'  molecular weight varies in a wide range (1-8 kDa) thereby resulting in bilayers with greater range of thicknesses, unlike lipids. The diblock polymers showed membrane viscosity spanning four orders of magnitude range,  from 14 nPa.s.m to 157000 nPa.s.m. The  lowest molecular-weight polymer membrane exhibits a viscosity similar to that of POPC.
 %thereby offering an opportunity  to test the widely assumed relation between 2D and 3D viscosity, $\eta_{2D}=h \eta_{3D}$. 
The membrane viscosity in Fig. \ref{polymer}A  does follow a power-law dependence on $M_h$, but the power 5.6 is much larger  than the expected range 0.5-1. In contrast, as seen from Fig. \ref{polymer}B, the bending rigidity follows a power-law  consistent  with the expected  $\kappa \sim M_h^{2n}$, since $\kappa/K_A\sim h^2$, where $K_A$ is the area-compressibility modulus \cite{Boal}.  The $K_A$ value is relatively insensitive to the molecular weight;  its value for the lipids was taken to be 250 mN/m and for the polymer membranes - 100 mN/m \cite{Rawicz:2000,Bermudez:2002}.
%While $K_A$ is likely to be independent of $M_h$, 
The polymer 3D viscosity $\eta$, however,  varies with the polymer molecular weight and this may be the  source of the unexpectedly higher exponent in  the power-law dependence of $\eta_\mem$ on $M_h$.
%As a result, the dependence of $\eta_\mem$ on $M_w$ becomes nontrivial.
Comparison of the dependence on the strain-rate of the 2D and 3D viscosities (Fig. \ref{polymer}C and Appendix Section S4) shows that in three dimensions the shear-thinning is more pronounced. This suggests that the polymer confinement into a bilayer impedes the microstructure dynamics responsible for the shear thinning. Intriguingly, taking the ratio of the 2D and 3D viscosities yields a thickness $h $ ranging from 200 to 350 nm
 %, depending on strain rate, 
much  larger than typical thickness measured with cryo-TEM \cite{Discher:1999}. 
%{\col{(INTERESTING: BOTH SD AND THIST TEST SUGGEST THAT OUR 2D VISCOSITY IS 10 TIMES LARGER)}} 
The disparity might arise from 
%the fact the our model neglects
 the thermally-driven membrane undulations, which result in interface roughness that  is in the order of 100 nm for a tension-free membrane and 20 $k_BT$ bending modulus. Thus the effective membrane thickness would be larger for 'fuzzy' membranes.

\section*{Discussion and Conclusions}

%We developed a facile method to quantify membrane fluidity of biomimetic membranes. 
We developed a method to measure the  shear viscosity of the two-dimensional bilayer  fluid  using the transient deformation of  a giant vesicle.  The approach is inspired by the interfacial rheology measurements using deformation of droplets and capsules \cite{Fischer:2007}.
 %  The method 
%is able to interrogate a wide range of bilayer compositions and phase states.
%, and the range of measured apparent membrane viscosity is more than four orders of magnitude.} 
% is sensitive to variations in the degree of unsaturation in the carbon chain of lipids in single-component lipid membranes, composition and phase state in multi-component membranes, and molecular weight  in block-copolymer membranes. The range of measured viscosities spans four orders of magnitude. 
 %The results are in good agreement with previous studies.
 The ``vesicle-rheometer'' enabled  the collection of an unprecedented amount of data, which provided valuable insights into the rheology of bilayer membranes and   lead to two new intriguing findings that raise more questions.
 % and need further systematic exploration. 
 
 First, bilayer membranes can behave as non-Newtonian fluids.
 % and the underlying mechanisms need further systematic exploration.
 Phase-separated bilayers are the 2D analogs of 3D dispersed systems such as emulsions and suspensions, which exhibit shear thinning behavior due to microstructure reorganization and relaxation; the domain dynamics in the phase-separated bilayers is likely the source of the shear thinning dynamics. Likewise, the shear thinning viscosity of  the diblock-copolymer membranes likely originates from polymer stretching and relaxation on a time scale comparable to the shearing. 
 
  Second, membrane viscosity  increases with field frequency reaching a plateau above few kHz. The zero-frequency limit of the viscosity is comparable to the one measured in the absence of electric field by other methods, thus the frequency effect is not related to the electric field amplitude. We hypothesize that the phenomenon is due to membrane electrical polarization.
  %due to the capacitive nature of the bilayer.  
 %The bilayer is impermeable to ions.  
 In an applied  electric field, ions brought by conduction accumulate at the membrane surfaces, because the bilayer is impermeable to ions (Fig. \ref{figQ}). 
   \begin{figure}[h]
\includegraphics[width=\columnwidth]{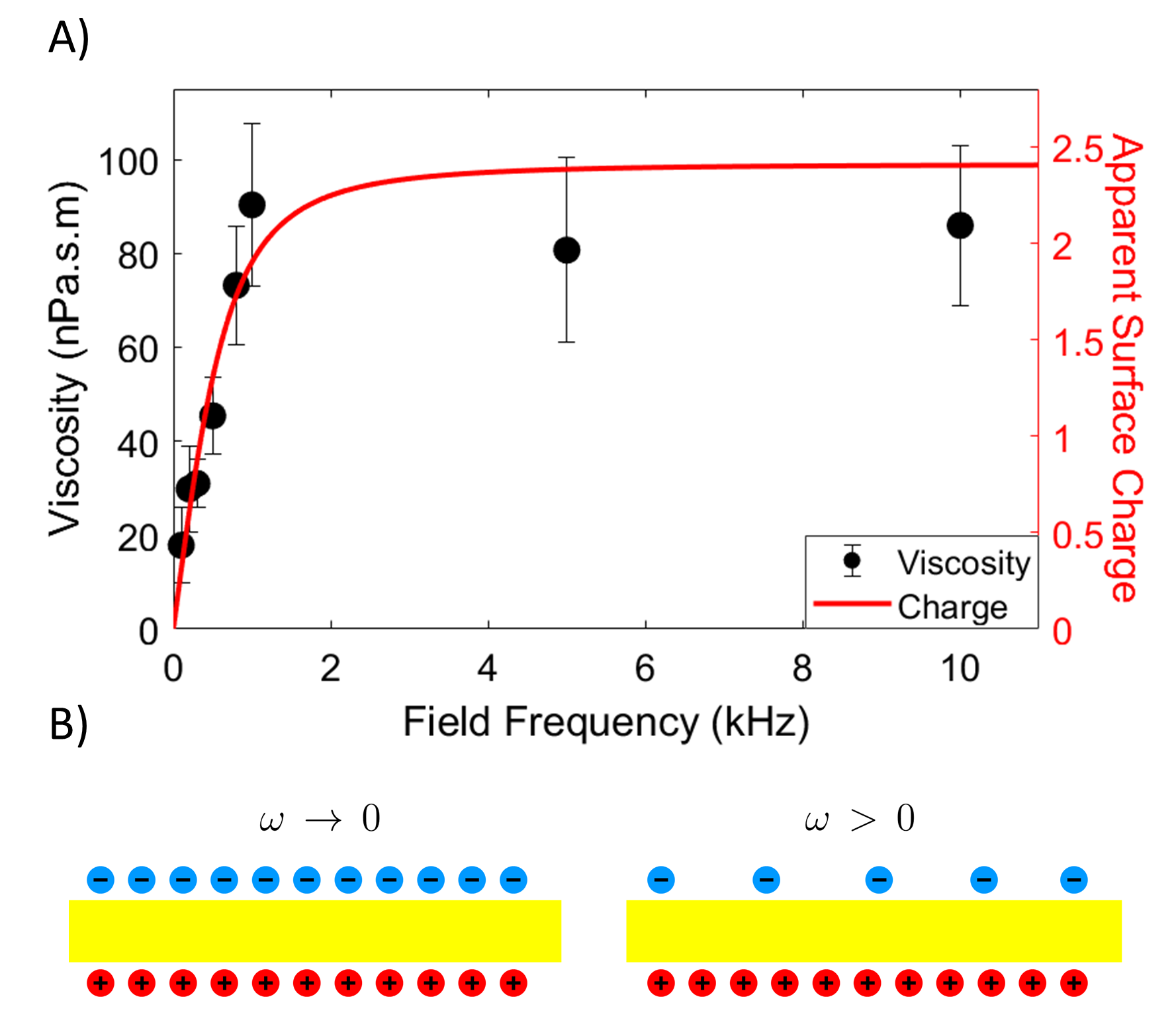}
\centering
\caption{ A) Membrane viscosity increases with field frequency. Data for SOPC, $E_0=8$ kV/m. Same behavior is exhibited by the imbalance in the surface charges accumulated at the opposing membrane surfaces computed from Eq. 7 in SI. B) Schematic illustration of the membrane (yellow) and free charges (blue and red) under an external electric field. The membrane capacitor is fully charged at low frequencies but becomes short-circuited at higher frequencies giving rise to an imbalance of the induced charge. }
%The maximum value of the apparent charge is $3(\Rr-1)/(Rr+2)$. }
\centering
\label{figQ}
\end{figure}
 %result in induced  surface charge of opposite sign (Fig. \ref{figQ).  
 At low frequencies, $\omega\ll\omega_c \sim \lambda_\ins /RC_m$, where $C_m$ is the membrane capacitance and $\lambda_\ins$ is the interior solution conductivity,  the membrane capacitor is fully charged and the net charge of the membrane (sum of the accumulated charge on the two surfaces) is zero.  As the frequency increases, $\omega\gtrsim\omega_c$,  the capacitor becomes short-circuited and draws current. This gives rise to an imbalance in the induced surface charge on the the opposite membrane surfaces; this imbalance (equivalent to an apparent membrane charge) correlates  remarkably well with the viscosity frequency dependence, see Fig. \ref{figQ}. The only study of  the effect of 
 %Increased 
  surface charge on the viscosity of bilayers, using mixtures of charged and neutral surfactants, has reported a significant increase  \cite{Bradbury}.  The bilayers in our study  are made of charge-neutral lipids, however, an apparent surface charge  originates from excess mobile charges (ions) at the membrane surfaces.
 %Coincidentally, the same behavior is exhibited by the induced charge on the membrane, see Fig. \ref{figQ}. 
 %Surface charges  been found to significantly increase viscosity of bilayers \cite{Bradbury}
The  mechanisms underlying the viscosity dependence on strain-rate and field frequency need further systematic exploration.

  The ease of implementation, high-throughput, minimal experimental equipment and effort as well as robustness  make this technique easy to adopt in every lab. The method can also be applied to obtain interfacial properties of lipid monolayers using deformation of droplets. 
  %Currently, viscosity of membranes containing charged species can only be extracted from the shape relaxation after the electric field is turned off. However, the approach using the initial slope of the vesicle deformation could be  implemented in a microfluidic analogue of the four-roll mill  or optical stretcher.  
  We envision that our approach will become a standard tool for characterization of membrane fluidity that will help address questions of biological and engineering importance such as mechanics of excitable cells and synthetic cell design. 

\section*{Materials and Methods}

\subsection*{Vesicle Preparation}
Giant unilamellar vesicles (GUVs) are formed from lipids and polymer such as palmitoyloleoylphosphatidylcholine (POPC), dioleoylphosphatidylcholine (DOPC), oleoylmyristoylphosphatidylcholine (OMPC), cholesterol (Chol),  stearoyloleoylphosphatdylcholine (SOPC), dipalmitoylphosphatidylcholine (DPPC) and poly(butadiene)-$b$-poly(ethylene oxide) diblock copolymers, PBd$_{x}$-$b$-PEO$_{y}$. The lipids and diblock copolymer were purchased from Avanti Polar Lipids (Alabaster, AL) and Polymer Source Inc. (Montreal, Canada), respectively. The multi-component vesicles made of DOPC:DPPC:Chol were fluorescently marked with 0.1 mol$\%$ of Liss Rhod PE. The lipid vesicles were produced using the electroformation method \cite{angelova.1987}. The stock solutions of 12 mM lipid in choloroform are diluted to 5 mM from which 10 $\mu$l of the solution is spread on the conductive sides of the ITO slides (Delta technologies, USA). The slides are stored in vacuum for 2-4 hours to evaporate all the organic solvents. The two slides are then sandwiched with a 2 mm thick teflon spacer and the electroformation chamber is filled with 40 mM sucrose solution in 0.3 mM of NaCl. The chamber is connected to a signal generator (Agilent, USA) for 2 hours at 50 Hz and voltage 1.5 V at 60$^o$ C, which ensures that all lipids are above their main phase transition temperatures. The harvested vesicles are diluted in isotonic glucose solution without salt. 3 independent GUV batches for every lipid composition were analyzed. Polymer vesicles were produced from spontaneous swelling method. Initially, 50 $\mu$l of 6-10 mg/ml (in chloroform) polymer solution was dissolved in 200-300 $\mu$l of chloroform in a 20 ml vial. Polymer films were formed from evaporation by blowing with a nitrogen stream while swirling the solution inside. Afterwards, the vials were dried under vacuum for 2-4 hours. The polymer films were hydrated in the suspending solutions (40 mM sucrose solution in 0.3 mM NaCl) and placed at 60 °C in an oven for 18-24 hours.

\subsection*{Electrodeformation}
The electrodeformation experiments are conducted in the electrofusion chamber (Eppendorf, Germany). The chamber is made from Teflon with two 92 $\mu$m cylindrical parallel electrodes 500 $\mu$m apart. The field is applied using a function generator (Agilent 3320A, USA). The function generator is controlled using a custom built MATLAB (Mathworks, USA) progam. This gives a precise control over the strength and duration of applied electric fields \cite{faizi2021}.
\subsection*{Optical microscopy and imaging}
The vesicles are visualized using a phase contrast microscope (A1 Axio Observer, Zeiss, Germany) with 63x objective 0.75 NA (air). Imaging is performed using Photron SA1.1 camera. The image acquisition rate for electrodeformation recordings is kept to a constant of 500-2000 fps for lipid vesicles and 60-200 fps for polymer vesicles and the shutter speed is fixed to 500 $\mu$s. The time evolution of the vesicle is analyzed using a home-made image analysis software. The software uses a Fourier series to fit around the vesicle contour, $r_s=\sum_{n=0}^\infty \alpha_n\cos(n \theta) +\beta_n \sin(n \theta)$. The second mode in the series is used to determine the major $(a)$ and minor axis $(b)$ of the deformed vesicles to evaluate $\nu=\frac{a}{b}={(1+\alpha_2)}/{(1-\alpha_2)}$. 
The confocal imaging was performed with Leica TCS SP8 scanning confocal microscope using a HC PL APO 40x/ NA 1.3 (oil) objective. The pinhole size during the experiment was fixed to 1 AU (Airy units). The dye was excited with a 561 nm laser (diode-pumped solid-state laser) with 1.61\% (laser intensity)  HyD3 detector (hybrid).
% and the gain was fixed to 23$\%$
\subsection*{Bending rigidity and tension measurements}
Flickering spectroscopy is a popular technique to extract out membrane rigidity and tension due to its non-intrusive nature and well developed statistical analysis criteria. The details of the technique are given in %\textit{Gracia et al.} and \textit{Faizi et al.}
 \cite{Gracia:2010,faizi_sm_2019, Faizi:2020}. Essentially, a time series of fluctuating vesicle contours is recorded on the focal plane. The quasi-circular  contour is represented in Fourier modes, $r(\phi)=R\left(1+\sum_q u_q(t) \exp(\im q \phi)\right)$. The fluctuating amplitudes $u_q$ 
have mean square amplitude  dependence on the membrane bending rigidity $\kappa$ and the tension $\sigma$, 
%\begin{equation}
$\langle \left|u_{q}\right |^2\rangle\sim\frac{\kT}{\kappa \left(q^3+\bar \sigma q\right)}$,
%\label{spectraeq}
%\end{equation}
where $\kT$ is the thermal energy (k$_B$ is the Boltzmann constant and T is the temperature), and $\bar\sigma=\sigma R^2/\kappa$. 
% The fluctuations were recorded with phase contrast microscope (Axio Observer A1  Zeiss, Germany) using a 63x/ Numerical Aperture (NA) 0.75 Ph2 (air) objective at 60 fps with high speed camera (Photron SA1.1). The Focal depth, $FD$, was determined using the standard formula $d=\frac{l}{NA^2}$ where the wavelength, $l$,  of transmission light is 550 nm. This results in FD= 0.97 $\mu{m}$ with the non dimensionalized $\Delta=\frac{FD}{R}$ (vesicle size range of radius, $R$, 20-50 $\mu{m}$) smaller than 0.05 to avoid the averaging effect of out of focus optical projections on equatorial projections \cite{Faizi:2020}. 
The integration time effect of the camera was reduced by acquiring images at a low shutter speed of 100-200 $\mu$s. At least 5000 images were obtained for each vesicle for good statistics.

\section*{Acknowledgmens}
P.M.V and H.A.F acknowledge financial support by NIGMS award 1R01GM140461.  This research was also supported in part by the National Science Foundation under Grant NSF PHY-1748958. }
%We thank Talbia Bint Humayun for proofreading the paper.}

\clearpage

\appendix
\begin{widetext}

\section{Compendium of published values for membrane viscosity}
see Table II

\begin{table*}[h]
\scriptsize
  \caption{\scriptsize{Membrane viscosity obtained using different experimental methods and simulations.* denotes membrane viscosity obtained from the bulk viscosity, $\eta$, using the relation $\eta_m=\eta{h}$, where $h$ is the bilayer thickness. For DMPC, SOPC and DOPC, the bilayer thickness are 3.67 nm, 4.00 nm \cite{KUCERKA20112761} and 3.67 nm \cite{pan2008} respectively. $L_d$ and $L_o$ refer to liquid-disordered and liquid-ordered phases respectively. GUVs, LUVs and SUVs refer to Giant-Unilamellar vesicles, Large-Unilamellar vesicles and Small-Unilamellar vesicles respectively. Zero-frequency membrane viscosities  obtained from the electrodeformation method developed in this study were measured at $E_0=8$ kV/m ($\dot{\gamma\,=\,50\,s^{-1}}$).}}
  \label{diffcondmethod}
  \begin{tabular*}{\textwidth}{@{\extracolsep{\fill}}lllll}
    \hline
    Method & Membrane Composition & Viscosity (nPa.s.m) & Temperature &  System\\
    \hline
    Falling Ball viscosimetry
    \cite{Dimova:1999}& SOPC& $3$ & 298 K& GUVs\\
     Coarse-grained simulations
    \cite{Shkulipa:2007}& SOPC& $3$ & 323 K& Bilayer\\
     Electrodeformation (this study)
     & SOPC & $9.73\pm5.80$ & 298 K& GUVs\\
    Probe Diffusion
    \cite{Gambin2098}& SOPC& $13.2\,^*$ & 293 K& GUVs\\
    Optical Dynamometry
    \cite{Dimova2000}& DMPC& $5\pm2$& 298 K& GUVs\\
    Neutron Spin Echo
    \cite{Nagao:2017}& DMPC& $75$  & 298 K& SUVs\\
    Coarse-grained simulations
    \cite{Shkulipa:2007}& DMPC& $5$ & 323 K& Bilayer\\
    Fluorescence Spectroscopy
    \cite{nojima2014}& DMPC& $0.36\,^*$  & 298 K& SUVs\\
    Tether pulling
    \cite{WAUGH198229}& EPC& $2.7-88$ & 295-298 K& GUVs\\
    Shear surface rheology
    \cite{Espinosa6008}& POPC& $300000$& NA& Monolayer\\
    Electrodeformation (this study)
     & POPC & $9.32\pm5.95$ & 298 K& GUVs\\
           Shear surface rheology
    \cite{Espinosa6008}& POPC:Chol (7:3)& $10000$ & NA& Monolayer\\
       Shear surface rheology
    \cite{Espinosa6008}& DPPC& $900000$ & NA& Monolayer\\
    Optical Tweezers
     \cite{Amadore2100156118}& DOPC & $<$0.6   & 293 K& Bilayer\\
     Fluorescence Spectroscopy
    \cite{nojima2014}& DOPC& $0.20\,^*$ & 298 K& SUVs\\
    Fluorescence Quantum Yield
    \cite{Wu2013}& DOPC& $0.84\,^*$ & 298 K& LUVs\\
    Fluorescence lifetime of dye
    \cite{chwastek2019}& DOPC& $0.15\,^*$ & 298 K& LUVs\\
     All-atom simulations 
    \cite{Zgosrki:2019}& DOPC& $0.197\pm0.0069$  & 297 K& Bilayer\\
    Coarse-grained simulations
    \cite{Shkulipa:2007}& DOPC& $0.17$ & 323 K& Bilayer\\
    Membrane-anchored particles
    \cite{Hormel:2014}& DOPC& $15.3\pm3.4$ & 297 K& GUVs\\
     Electrodeformation (this study)
     & DOPC& $4.11\pm2.63$ & 298 K& GUVs\\
     Probe Diffusion
     \cite{Herold:2013}& DOPC& $0.59\pm0.2$ & 298 K& GUVs\\
      Neutron Spin Echo
     \cite{Chakraborty21896}& DOPC& $16.7\pm1.1$  & 298 K& SUVs\\
          Electrodeformation (this study)
     & OMPC& $7.73\pm3.09$ & 298 K& GUVs\\
     Membrane-anchored particles
     \cite{Hormel:2014}& 13:0 PC& $14.7\pm6.9$  & 297 K& GUVs\\    Neutron Spin Echo
    \cite{Chakraborty21896}&
    DOPC:Chol (8:2)& $31.9\pm3.5$ & 298 K& SUVs\\
    Electrodeformation (this study)
     &DOPC:Chol (1:1) & $7.00\pm4.77$ & 298 K& GUVs\\
    Electrodeformation (this study)
     &DPPC:Chol (1:1)  & 56.4$\pm4.63$ & 298 K& GUVs\\
        Optical Tweezers
     \cite{Amadore2100156118}& DOPC:DPPC (2:1), $L_o$ & $2.1$   & 293 K& Bilayer\\
    Shear Driven Flow \cite{Honerkamp:2013}& DOPC:DPPC (85:15), $L_d$ & 1.9$\pm$11.9  & 296 K&GUVs\\
    Shear Driven Flow \cite{Honerkamp:2013}& DiPhyPC:Chol:DPPC (5:40:55), $L_o$& 15.7$\pm$9.9  &296 K& GUVs\\
    Domain fluctuations \cite{Camley:2010}& DiPhyPC:Chol:DPPC (25:55:20) & 4$\pm$1   & 293 K& GUVs\\Diffusion of domains
    \cite{Cicuta:2007}& DPPC:DOPC:Chol (3.5:3.5:3), $L_o$ & 10-500 & 295 K& GUVs\\
    Diffusion of domains
    \cite{Petrov:2012}&DiPhyPC:DPPC (1:1) & 2.2$\pm0.1$  & 296.5 K& GUVs\\
     Shear Driven Flow
     \cite{Imai:2020}& DPPC:DOPC:Chol (3:6:1),  $L_o$ domain& $4.1$ & 293 K& GUVs\\
     Shear Driven Flow
     \cite{Imai:2020}& DPPC:DOPC:Chol (4:5:1), $L_o$ domain& $14$  & 293 K& GUVs\\
     Shear Driven Flow
     \cite{Imai:2020}& DPPC:DOPC:Chol (2:6:2), $L_o$ domain& $9.6$  & 293 K& GUVs\\
    Shear Driven Flow
     \cite{Imai:2020}& DPPC:DOPC:Chol (4:4:2), $L_o$ domain& $4.3$    & 293 K& GUVs\\
     Shear Driven Flow
     \cite{Imai:2020}& DPPC:DOPC:Chol (5:3:2), $L_o$ domain& $54$  & 293 K& GUVs\\
     Shear Driven Flow
     \cite{Imai:2020}& DPPC:DOPC:Chol (6:2:2), $L_d$ domain& $1200$  & 293 K& GUVs\\
      Shear Driven Flow
     \cite{Imai:2020}& DPPC:DOPC:Chol (2:5:3), $L_o$ domain& $7.7$ & 293 K& GUVs\\
     Shear Driven Flow
     \cite{Imai:2020}& DPPC:DOPC:Chol (5:2:3), $L_d$ domain& $85$  & 293 K& GUVs\\
     Shear Driven Flow
     \cite{Imai:2020}& DPPC:DOPC:Chol (6:1:3), $L_d$ domain& $1200$  & 293 K& GUVs\\
     Shear Driven Flow
     \cite{Imai:2020}& DPPC:DOPC:Chol (3:3:4), $L_d$ domain& $0.63$ & 293 K& GUVs\\
     Shear Driven Flow
     \cite{Imai:2020}& DPPC:DOPC:Chol (4:2:4), $L_d$ domain& $65$   & 293 K& GUVs\\
     Shear Driven Flow
     \cite{Imai:2020}& DPPC:DOPC:Chol (5:1:4), $L_d$ domain& $71$   & 293 K& GUVs\\
     Shear Driven Flow
     \cite{Imai:2020}& DPPC:DOPC:Chol (4:3:3), $L_d$ domain& $98$   & 293 K& GUVs\\
     Shear Driven Flow
     \cite{Imai:2020}& DPPC:DOPC:Chol (3:5:2), $L_o$ domain& $65$   & 293 K& GUVs\\
    Electrodeformation (this study)
     &DPPC:DOPC:Chol (1:1:1) & $17.7\pm3.06$ & 298 K& GUVs\\
     Electrodeformation (this study)
     &DPPC:DOPC:Chol (1:1:2) & $15.4\pm2.40$ & 298 K& GUVs\\
     Falling Ball viscosimetry \cite{Dimova2002}
     & PBd$_{33}$-$b$-PEO$_{20}$ & $1500\pm120$ & 298 K& GUVs\\
    Electrodeformation (this study)
     & PBd$_{33}$-$b$-PEO$_{20}$ & $2890\pm670$ & 298 K& GUVs\\
    Micropipette Aspiration \cite{Mabrouk2009}
     & PEG-b-PA6ester1 & $7900000\pm200000$ & 298 K& GUVs\\
    Micropipette Aspiration \cite{Mabrouk2009}
     & PEG-b-PA444 & $4000000\pm200000$ & 298 K& GUVs\\
         Atomic Force Microscopy \cite{janshoff2010}
     & PBd$_{130}$-$b$-PEO$_{66}$ & $5000000$ & 298 K& Bilayer\\
         Atomic Force Microscopy \cite{janshoff2010}
     & PBd$_{130}$-$b$-PEO$_{66}$ (cross linked) & $20000000$ & 298 K& Bilayer\\
     Electrodeformation (this study)
     & PBd$_{13}$-$b$-PEO$_{11}$ & $14.4\pm4.40$ & 298 K& GUVs\\
     Electrodeformation  (this study)
     & PBd$_{22}$-$b$-PEO$_{14}$ & $686\pm51.0$ & 298 K& GUVs\\
     Electrodeformation (this study)
     &PBd$_{46}$-$b$-PEO$_{24}$ & $20600\pm4700$ & 298 K& GUVs\\
      Electrodeformation (this study)
     &PBd$_{54}$-$b$-PEO$_{29}$ & $46700\pm900$ & 298 K& GUVs\\
      Electrodeformation (this study)
     &PBd$_{120}$-$b$-PEO$_{78}$ & $157000\pm54500$ & 298 K& GUVs\\
    \hline
    \centering
  \end{tabular*}
\end{table*}

\newpage

\section{Shape evolution of quasi-spherical vesicle in a uniform electric field}
\label{theory}

\subsection{Theoretical model}
%\begin{wrapfigure}[12]{r}{2.in}
%\centerline{\includegraphics[width=1.5in]{figS3.png}}
%\end{wrapfigure}
\begin{figure}[h]
\includegraphics[width=\columnwidth]{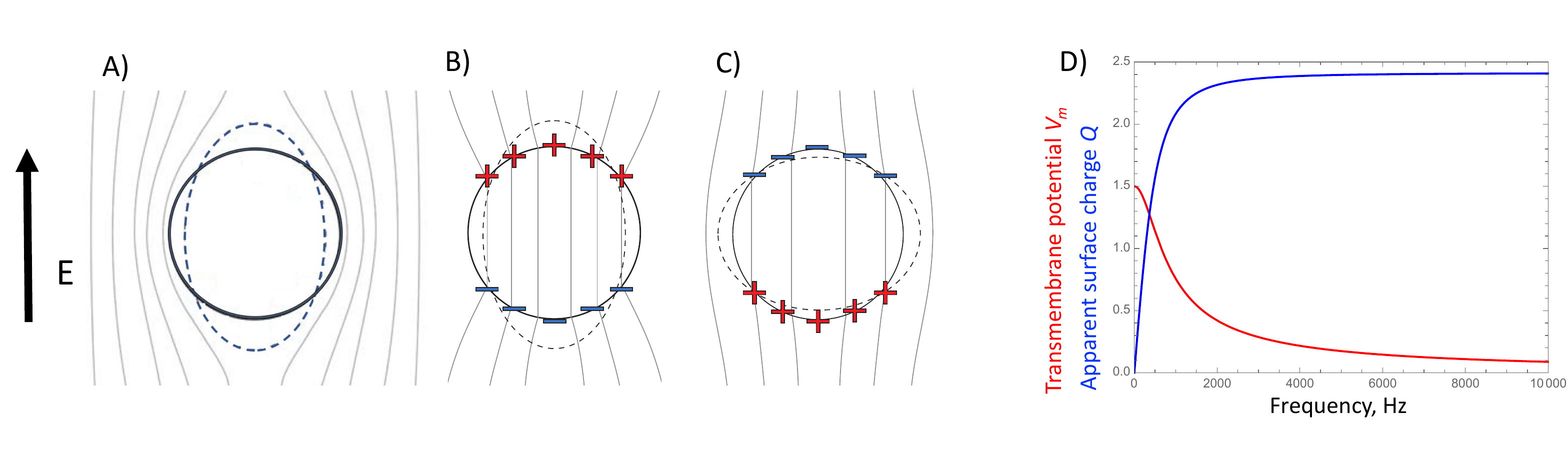}
\caption{ (A)-(C) Physical mechanisms of the frequency-dependent membrane polarization and vesicle dipole
 in an applied   uniform AC field.  The lines correspond to constant electric field. Upon application of an external electric field, charges accumulate on the two sides of the  membrane setting up a potential difference, i.e., the  membrane acts as a capacitor.   (A)  At low frequencies, $\omega\ll\omega_c$, the membrane capacitor is fully charged, the induced charge density on the two membrane surfaces  is the same but of opposite sign. (B) and (C) At intermediate  frequencies, $\omega>\omega_c$,  it is short-circuited and there is charge imbalance between the inner and outer membrane surfaces $Q=\eps E_0 Q_0\cos\theta$.  (B) If the enclosed solution is more conducting than the suspending medium, $\Rr>1$, vesicle is pulled into an prolate ellipsoid.
(C)  The polarization is reversed in the opposite case  $\Rr<1$ and the vesicle deforms into an oblate ellipsoid. (D) Variation with frequency of the transmembrane potential (red) and apparent charge at the pole (blue).}
\label{figF}
\end{figure}

%Upon application of an external electric field, charges accumulate on the two sides of the  membrane setting up a potential difference, i.e., the  membrane acts as a capacitor.  If the capacitor is fully charged, it draws zero current and the voltage drop occurs entirely across the membrane. 

%basis of the electrodeformation method : vesicle shape in a uniform electric field}
Let us consider a vesicle made of a
charge-free lipid bilayer  membrane with  bending rigidity $\kappa$, tension $\sigma_\eq$, capacitance $C_\mem$. 
The vesicle is suspended in a
solution with conductivity $\sigm_\out$ and permittivity $\eps_\out$, and filled with a different
solution characterized by $\sigm_\ins$ and $\eps_\ins$.

An axisymmetric stress, such as generated by uniform electric field or extensional flow, 
%A uniform AC electric field with amplitude $E_0$ and frequency $\omega$, $\bE=E_0\cos(\omega t)\zhat$, 
deforms the vesicle into a spheroid
%exerts electric stress on the vesicle and causes it to deform into an
 with symmetry axis aligned with the extensional axis. The spheroid aspect ratio is $\nu=a/b$, where $a$ is the length of the symmetry axis and $b$ is the length of the axis perpendicular to the symmetry axis.
 For small deformations, $\nu\lesssim 1.3$, the shape is well approximated by  
\begin{equation}
r_s(\theta)=R\left(1+\frac{s}{2}\left(1+3 \cos2\theta\right)\right)\,,
%\quad \nu=1+3s
\end{equation}
where $r_s$ is the position of the surface,   $R$ is the initial radius of the vesicle, $s$ is the deformation parameter, and $\theta$ is the angle with the applied field direction; $\theta=0$ and $\pi/2$ correspond to the pole and the equator, respectively. The ellipsoid aspect ratio  is related to the deformation parameter by  $\nu=(1+s)/(1-2s)$.

The theory developed by \textit{Vlahovska et al.} \cite{Vlahovska-Dimova:2009,Vlahovska:APBL,Vlahovska:2019,Vlahovska:2019book} predicts that the  deformation parameter evolution is given by the balance of imposed and membrane stresses
\begin{equation}
\label{eqS1}
\dot s=  \frac{1}{32+23\visrat+16\visrat_m}\left(\frac{ \eps_\out E_0^2 p^\el}{\eta} -24 s(6\kappa +\sigma (s) R^2)\frac{1}{\eta R^3}\right)
%\eps_\out E_0^2 R^3 \frac{ p^\el (\omega)}{24(6\kappa +\sigma R^2)} \,.
%\frac{\eps E_0^2 R^3 p(\omega, \Rr, \Sr)}{\kappa\left(6+\bar \sigma\right)} 
\end{equation}
%\begin{equation}
%c=  \frac{ \eps_\out E_0^2}{\eta} p^\el (\omega)
%\end{equation}
%The membrane tension dependence on the vesicle deformation is given by
%\begin{equation}
%\sigma(s)=\sigma_\eq\exp\left(\frac{64\pi \kappa}{5\kT}s^2\right)
%\end{equation}
The AC field, $E(t)=E_0 \sin(\omega t)$, generates an electric stress which has two components, a steady one $p^\el$  and an oscillatory one with frequency twice the applied one
\[
p=p^\el+p^c_\omega\cos(2 \omega t)+p^s_\omega\sin(2 \omega t)
\]
In the experiments typically $\bar\omega\gg 1$ and the oscillatory component only drives very small oscillations about the deformation induced by the steady stress component. 

The steady electric stress is given by
 \begin{equation}
\label{elstress}
\begin{split}
p^\el= 
2(1- P^r_\out)+\frac{1}{2} P_\out^2-2 \Sr P_\ins^2
%\left(2(1-\tilde P_\out)+5P_\out^2-2\Sr P_\ins^2\right)
  \end{split}
\end{equation}
and the amplitudes of the unsteady stress are
\begin{equation}
\label{elstressP}
\begin{split}
p^c_\omega &=\frac{1}{2}\left(4(1- P^r_\out)-\left(P^i_\out\right)^2+\left(P^r_\out\right)^2-4\Sr\left(\left(P^i_\ins\right)^2+\left(P^r_\ins\right)^2\right)\right)\\
p_\omega^s&=2 P^i_\out-P^i_\out P^r_\out+4 \Sr P^i_\ins P^r_\ins
 \end{split}
\end{equation}
where 
 \begin{equation}
 \begin{split}
 P_\out=&\frac{K_\out+K_\ins(V_m-1)}{K_\ins+2K_\out}\,,\quad P_\ins=\frac{K_\out(3-2V_m)}{K_\ins+2K_\out}\,,\\
 V_m=&\frac{3 K_\ins K_\out}{2K_\ins K_\out+\im C_\mem\left(K_\ins+2 K_\out\right)\bar\omega}
 \end{split}
\end{equation}
Here $\bar\omega=\omega \eps_\out/\sigm_\out$ and $\Cm_\mem=C_\mem R/\eps_\out$ are the dimensionless frequency and membrane capacitance. $K_\ins=1+\im \bar\omega$ and $K_\out=\Rr+\im \bar\omega \Sr$ are the dimensionless complex permittivities. 
$\Sr=\eps_\ins/\eps_\out$ and $\Rr=\sigm_\ins/\sigm_\out$ are the ratios of permittivities and conductivities of the fluids interior and exterior to the vesicle.  $P^r$ and $P^i$ denote the real and imaginary part of $P$, and $P^2=PP^*$, where the superscript * denotes complex conjugate. The electric stress in DC field is obtained by setting $\bar\omega=0$ and the electric field amplitude to $E_0\sqrt{2}$.

Typically, both the inner and outer fluids are aqueous solutions with similar permittivities, $\eps_\ins\approx \eps_\out=\eps$, hence $\Sr$ can be set to 1. In this case \refeq{elstress}  reduces to
\begin{equation}
\label{eqP}
\begin{split}
%\frac{\dif s}{\dif t}=\frac{p-24\left(\tilde\sigma +6 \tilde\kappa\right)s(t) }{\eta(32+23\visrat+16\visrat_m)} \,,\quad 
%s\left(\bar\omega\right)=
p^\el= \frac{9 \left[\bar\omega^2 \left(\Cm_\mem^2 (\Rr+2)^2 \left(\Rr-1\right)\left(\Rr+3\right)+2  \Cm_\mem\Rr \left(\Rr^2+\Rr-2\right)+9 \Rr^2\right)+\Rr^2 (\Rr+2)^2\right]}{2 \left((\Rr+2)^2+9 \bar\omega^2\right) \left(\Cm_\mem^2 (\Rr+2)^2 \bar\omega^2+4 \Rr^2\right)}\,,
\end{split}
\end{equation}
where $\bar\sigma=\sigma R^2/\kappa$. At low frequencies, $\bar\omega\rightarrow 0$, the membrane capacitor is fully charged, and $p^\el=9/16$ and we obtain Equation  1 in the main text.

The imbalance between the induced charge of the two membrane surfaces is $Q=\eps E_0 Q_0\cos\theta$, where the maximum charge is 
\begin{equation}
Q_0=\frac{3\bar \omega C_m\left(\Rr-1\right)\left(\Rr+2\right)}{\left[\left((2+\Rr)^2+9\bar\omega^2\right)\left(4\Rr^2+C_m^2\bar\omega^2(2+\Rr)^2\right)\right]^{1/2}}
\end{equation}
At low frequencies, $\bar\omega\rightarrow 0$ the charge imbalance vanishes. 

\subsection{Linear approximation of the evolution equation }
Assuming constant tension, Eq. 1 in the main text can be integrated to yield
\begin{equation}
\label{shapeevSM}
\begin{split}
\nu(t)=1+\frac{p R}{24\sigma}\left(1-\exp{\left(-\frac{24\sigma}{\eta R\left(55+16\visrat_m\right)}{t}\right)}\right)
%\nu(t)=1+\frac{9\eps E_0^2 R}{64\sigma}\left(1-\exp{\left(-\frac{24\sigma}{\eta R\left(55+16\visrat_m\right)}{t}\right)}\right)
\end{split}
\end{equation}
If 
\[
\frac{24\sigma t}{\eta R\left(55+16\visrat_m\right)}\ll 1
\]
the exponent is expanded in Taylor series to yield the linear evolution, Eq. 2 in the main text.

The time limit for the linear approximation can be estimated  by comparing the linear and quadratic terms in the Taylor series of the exponential term in \refeq{shapeevSM}.
% The two terms become equal if $t/t_\ehd=(55+16\visrat_m)\eps E_0^2/12\sigma$. 
%More precisely, the linear approximation is only valid if the argument of the exponential function is
The Taylor series of the exponential function is
\[
\exp\left(-c t\right)=1-c t+\frac{(ct)^2}{2}+h.o.t.\quad \mbox{where for \refeq{shapeev}}\quad c=\frac{24\sigma}{\eta R\left(55+16\visrat_m\right)}
\]
It shows that the quadratic correction becomes comparable to the linear term when
\[
ct=\frac{(ct)^2}{2}\Longrightarrow t=\frac{2}{c}
\]
which gives the estimate for the  time up to which the linear approximation is reasonable
\[
\frac{t}{t_\ehd}=\frac{\eps E_0^2 R\left(55+16\visrat_m\right)}{12\sigma}
\]
%Considering typical parameters  $\eps E_0^2/\sigma\sim1$, yields $t/t_\ehd\sim 5$ if $\visrat_m\sim 0$. Higher membrane viscosity extends the linear deformation regime, if $\visrat_m\sim 10$, as in polymersomes made of PBd$_{22}$-$b$-PEO$_{14}$, where the cut-off time becomes $t/t_\ehd\sim15 $, see Fig. 1D in the main text. 

\newpage

\section{Additional data}

 \subsection{Batch reproducibility for homogeneous (DOPC) and mixed membrane compositions (DOPC:DPPC:Chol (1:1:1))}
\begin{figure}[h]
\includegraphics[width=3in]{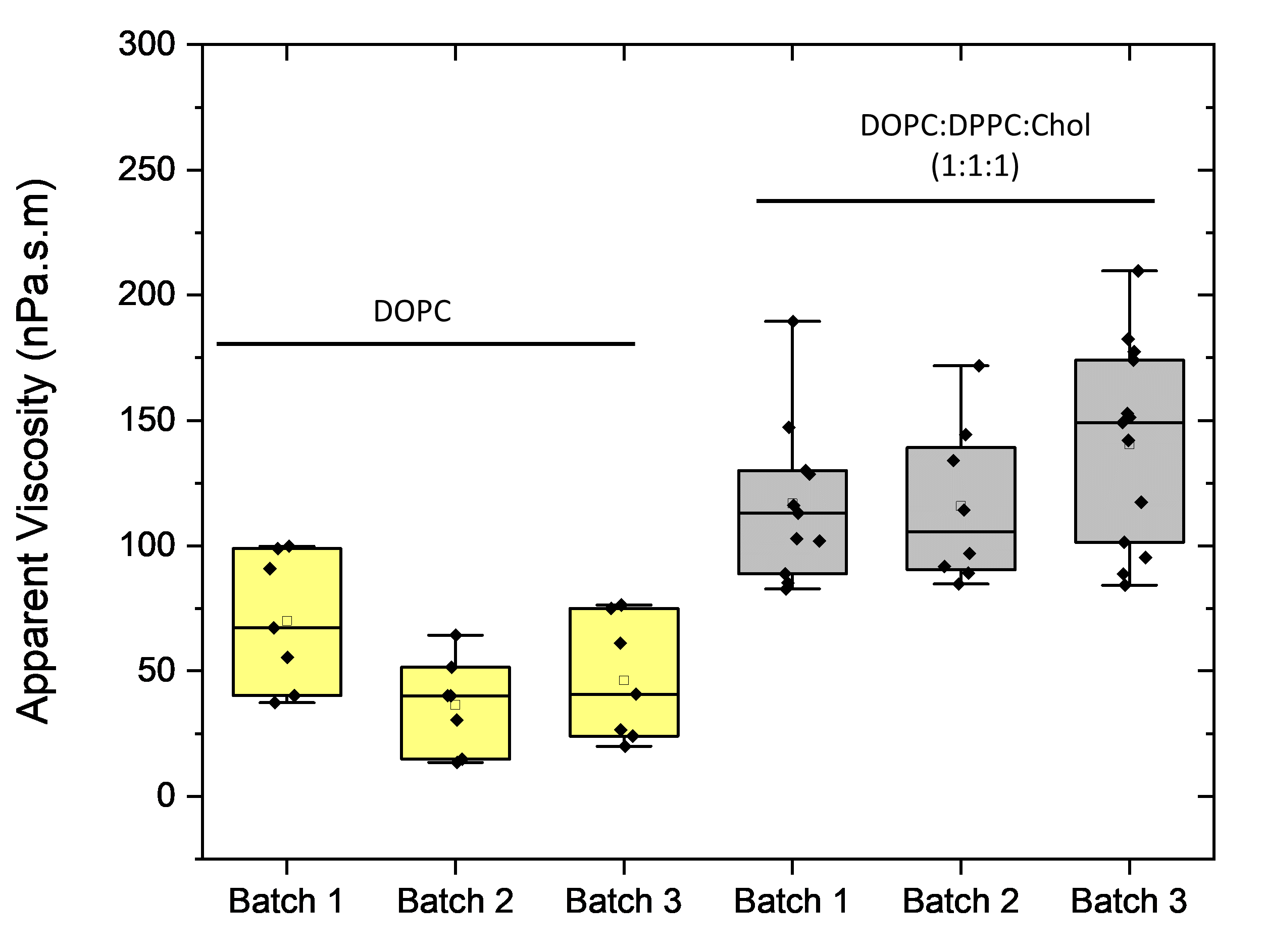}
\centering
\caption{Method reproducibility for the same system across different batches: Membrane viscosity for DOPC and DOPC:DPPC:Chol (1:1:1) for three different batches of prepared vesicles. The applied frequency and field strength are 1 kHz and 10 kV/m for respectively. For DOPC:DPPC:Chol (1:1:1), the applied frequency is 2 kHz and and applied field strength is 6 kV/m. The solid symbols show measurements on individual vesicles. The box-plot represents the standardized distribution of data based on five numbers minimum value, first quartile (Q1), median, third quartile (Q3), and maximum value. The open square represents the mean value..}
\centering
\label{fig_reproducible_batch}
\end{figure}

Figure \ref{fig_reproducible_batch} shows the box plot presentation for apparent membrane viscosity values obtained for the same system across three different batches of vesicles prepared from DOPC and DOPC:DPPC:Chol (1:1:1). The zero charge or frequency membrane viscosity data is given in the main text. 
 
%  \begin{table}[h]
% \caption{Method reproducibility for the same system across different batches: Average apparent membrane viscosity for SOPC, DOPC and DOPC:DPPC:Chol (1:1:1) for three different batches of prepared vesicles from data of Figure \ref{fig_reproducible_batch}. The applied frequency and field strength are 1 kHz and 10 kV/m for SOPC and DOPC respectively. For DOPC:DPPC:Chol (1:1:1), the applied frequency is 0.2 kHz and and applied field strength is 6 kV/m.}
% \centering
% %\begin{threeparttable}
% \begin{tabular}{|c|c|c|c|}
% \hline
% Composition& Batch 1 ($10^{-8}$ Pa.s.m)&Batch 2 ($10^{-8}$ Pa.s.m)&Batch 3 ($10^{-8}$ Pa.s.m) \\\hline\hline
% SOPC &$19.00\pm2.7$ &$23.19\pm3.16$ &$20.77\pm2.81$  \\\hline
% DOPC &$6.99\pm2.48$ & 3.64$\,\pm$2.02&$4.62\pm2.65$ \\\hline
% DOPC:DPPC:Chol (1:1:1)&$11.72\pm3.18$ & 11.58$\,\pm$3.29&$14.04\pm4.12$ \\\hline

% \end{tabular}
% \label{reproducibility}
% \end{table}

% \begin{figure}[h]
% \includegraphics[width=\columnwidth]{FigureS1.png}
% \centering
% \caption{a) The experimental setup b) Deformation of $R=$ 2 mm size water droplet in castor oil without any surfactants c) Deformation of $R=$ 2 mm size water droplet in castor oil with 0.5 $\%$ Tween 80. 
% }
% \centering
% \label{fig_setup}
% \end{figure}

 \subsection{Deformation curves of  bilayers at different field strength and frequency}
 Figure \ref{POPC_different_Field}A represents deformation curves of POPC vesicles at different field strength (6-10 kV/m) at  frequency 1 kHz. In Figure \ref{POPC_different_Field}B the data is re-plotted again in rescaled time with $t_\ehd$. The inverse of $t_\ehd$ can be expressed as shear rate, $\dot{\gamma}$ which in this case ranges from, $\dot{\gamma}$ $\sim$10-100 s$^{-1}$. The collapse of the data on single curve indicates that the deformation rate of POPC bilayers are not affected at a given shear rate and they exhibit Newtonian rheology.

\begin{figure}[h]
\includegraphics[width=\linewidth]{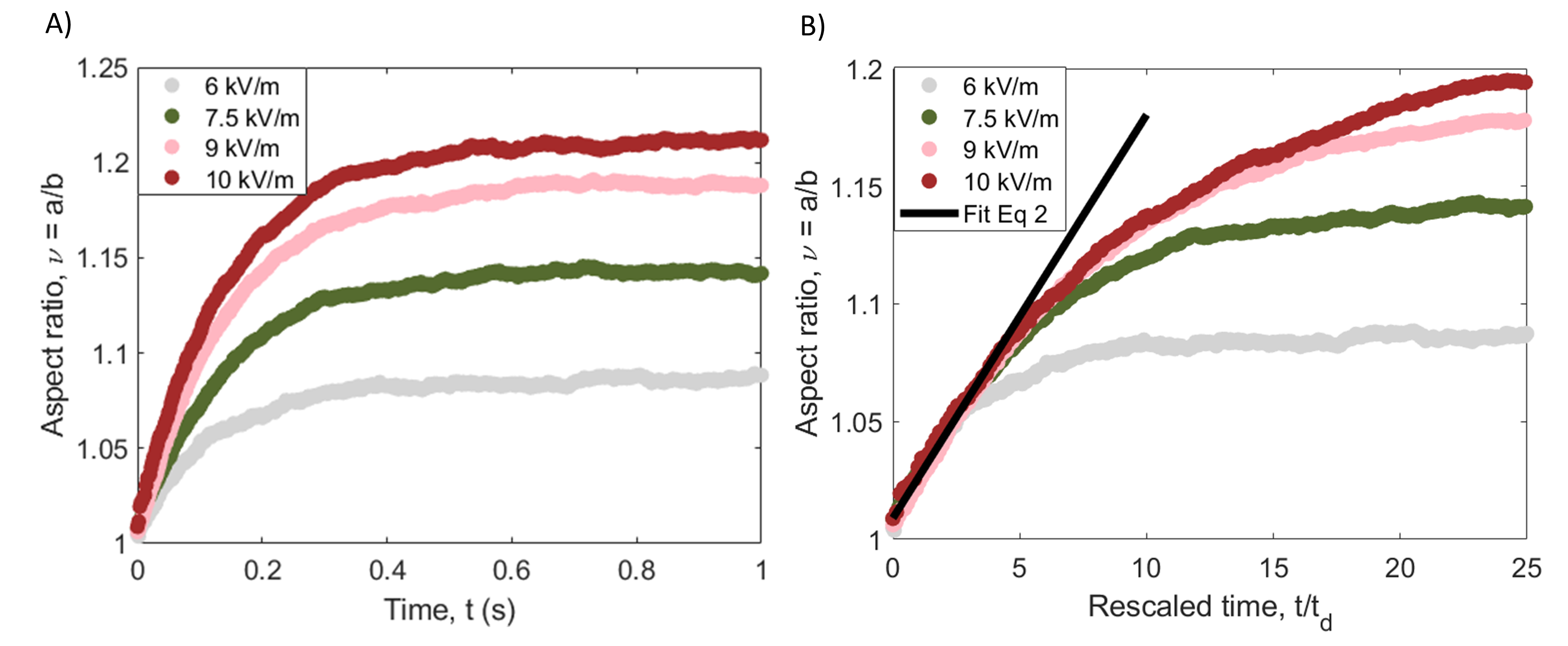}
\centering
\caption{(A) Deformation curves for a POPC vesicle (R= 30.1 $\mu$m) exposed to fields of different amplitudes (at 1 kHz).  (B) The initial slope of the data in (A) re-plotted as a function of the re-scaled time $t/t_\ehd$ yields an apparent membrane viscosity $\eta_m=2.63\pm0.41\times10^{-7}$ Pa.s.m.}
\centering
\label{POPC_different_Field}
\end{figure}

% \subsection{Deformation curves of bilayers at different frequency.}
 Eq. 2 in the main text shows that the slope depends on $t_\ehd$, which depends on the field amplitude $E_0$, and $p^\el$, which depends on frequency. Hence to isolate the viscosity, one needs to plot the deformation data as a function of time rescaled as $t/t_\ehd p^\el$, see Figure \ref{POPC_different_freq}.
\begin{figure}[h]
\includegraphics[width=\linewidth]{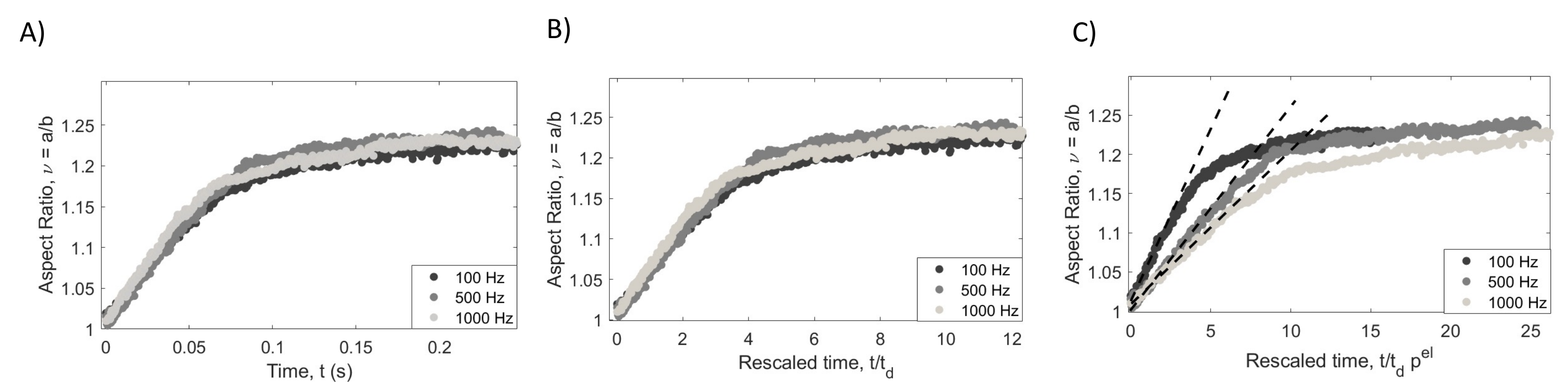}
\centering
\caption{(A) Deformation curves for a POPC vesicle (R= 14.7 $\mu$m) exposed to fields of different frequency but same field amplitude $E_0=8$ kV/m.  (B) The initial slope of the data in (A) re-plotted as a function of the re-scaled time $t/t_\ehd$, see Figure \ref{POPC_different_freq}.  The electric stress  $p^\el$ increases with frequency but the slope in (B) remains the same indicating that apparent surface viscosity also increases. (C) Indeed, when data are plotted vs $t/t_\ehd p^\el$ the slope  decreases with increasing frequency yielding the frequency dependence of the apparent viscosity. Extrapolation to zero frequency gives the membrane viscosity }
\centering
\label{POPC_different_freq}
\end{figure}

 \subsection{Newtonian and Non- Newtonian bilayers}
Figure \ref{flow curve} shows flow curve experiment (membrane viscosity vs imposed shear rate) for different compositions. Unsaturated phospholipids, for example DOPC in \ref{flow curve}A and B, 
are known to assemble into bilayers in the  liquid disordered, $L_d$ phase. Our data shows that at different shear rate, $\dot{\gamma}$ $\sim$10-1000 s$^{-1}$, or E- field ranging from 5-40 kV/m, the membrane viscosity of $L_d$ bilayers remains unchanged. Therefore, DOPC bilayers are 2D Newtonian behavior. In the next case, see \ref{flow curve}C, we analyze, fully saturated
lipids, DPPC, mixture with Cholesterol at 1:1 molar ratio. DPPC has a phase transition temperature temperature at 314 K. Adding Cholesterol to the bilayers results in the disruption of the gel or solid phase of the bilayer to $L_o$ state which has a larger extent of orientational order than $L_d$. Our results show that $L_o$ are also Newtonian in nature. Next we move to mixed systems exhibiting phase separation. 1:1 DOPC:DPPC demonstrated phase separation with intricate network of finger-like domains indicative of gel or solid phase. Interestingly, the flow curve experiments,\ref{flow curve}D and E, show Non- Newtonian characteristics with shear thinning behavior (undergoing flow with a decreasing viscosity with increasing shear rate). Polymer vesicles assembled from di-block copoylmers such as PBd$_{x}$-$b$-PEO$_y$ also show a shear thinning behaviour with increasing shear rate. \ref{flow curve}F and G gives two examples of PBd$_{x}$-$b$-PEO$_y$ bilayers where the higher $M_w$ polymersomes, PBd$_{33}$-$b$-PEO$_{20}$, show a much stronger shear thinning behaviour. The viscosity of PBd$_{33}$-$b$-PEO$_{20}$ decreases by a $100\%$ from $622\pm70\,\times\,10^{-8}$ Pa.s.m to $311\pm62\,\times\,10^{-8}$ Pa.s.m between shear rate $\dot{\gamma}$ $\sim$10-1000 s$^{-1}$. PBd$_{13}$-$b$-PEO$_{11}$ viscosity decreases from $8.72\pm2.22\,\times\,10^{-8}$ Pa.s.m to $5.6\pm0.5\,\times\,10^{-8}$ Pa.s.m between similar shear rates.

\begin{figure}[h]
\includegraphics[width=\textwidth]{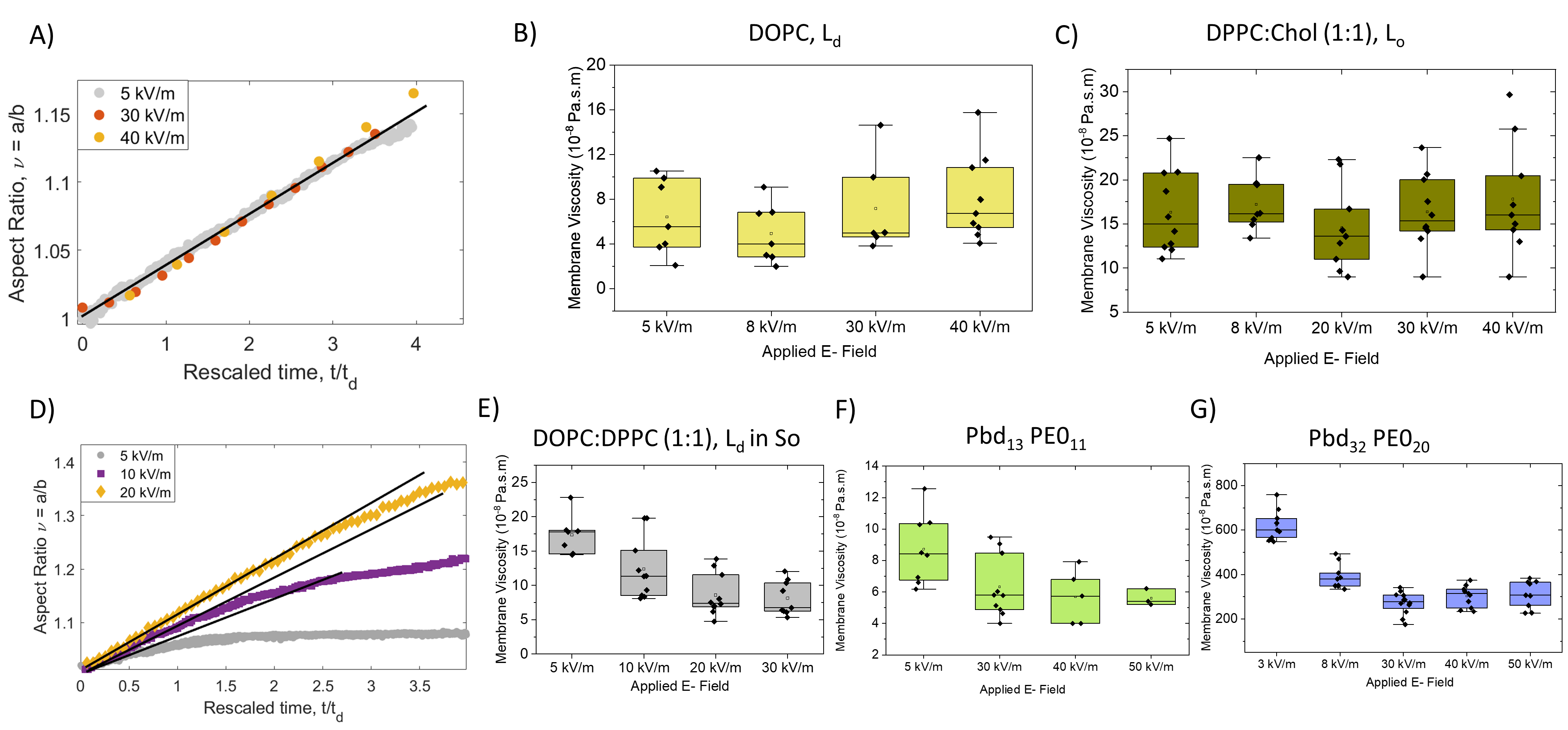}
\centering
\caption{Shear viscosity of different bilayer compositions at different electric field (3-50 kV/m) or shear rates $\dot{\gamma}\sim$ (1-1000) s$^{-1}$ at a fixed frequency 1 kHz. A) Deformation case of DOPC bilayer at different field strengths. The slope remains same regardless of the applied shear rate indicating 2D Newtonian fluids. B) Viscosity obtained at different field strengths for Newtonian cases B) DOPC C) DPPC:Chol (1:1) bilayers. D)  Deformation case of DOPC:DPPC (1:1) bilayer at different field strength. The slope increases as the field strength or shear rate is increased. This behavior is represented by shear thinning fluids. B) Viscosity obtained at different field strengths for non-Newtonian cases E) DOPC:DPPC (1:1) F) PBd$_{13}$-$b$-PEO$_{11}$ G) PBd$_{32}$-$b$-PEO$_{20}$.}
\centering
\label{flow curve}
\end{figure}

\subsection{Bending rigidity values from Flickering Spectroscopy and capacitance measurements for electrodeformation method}

The method for flickering spectroscopy is detailed in \cite{faizi_sm_2019, Faizi:2020}. Here, we summarize the electrodeformation method to extract out membrane capacitance. The procedure follows the original approach developed by \textit{Salipante et al.} \cite{Salipante-Vlahovska:2012}. The vesicle shape morphology with conductivity ratio $\Rr>1$ is always prolate. However, for $\Rr<1$, the conductivity of the outer solution is higher than the vesicle solution and the aspect ratio/deformation parameter $s\left(\omega\right)$ is positive at low frequencies that is prolate shape. As the frequency increases, the vesicles becomes less prolate and adopts a spherical shape at a certain frequency. Above this critical frequency, the vesicles adopt an oblate shape. 
The critical frequency can be approximated as:
\begin{equation}
\omega_c=\frac{\sigm_\ins}{R C_\mem}\frac{1}{\sqrt{\left(1-\Rr\right)\left(3+\Rr\right)}}
\end{equation}
 
Hence, the membrane capacitance can be determined from the experimentally measured critical frequency based on prolate-oblate transition with a frequency sweep \cite{Salipante-Vlahovska:2012}. The measured bending rigidity and capacitance values are summarized in Table \ref{polymerBR}.

\begin{table}[h]
\caption{Membrane bending rigidity and capacitance of  phospholipids, polymers PBd$_{x}$-$b$-PEO$_y$ and mixed system of DOPC:DPPC:Chol at 25 $^o$C determined in this study. Bending rigidity was measured with flickering spectroscopy and  membrane capacitance was measured with the electrodeformation method. $M_w$ and  $M_h$ refer to the total and hydrophobic molecular weight, respectively. NA means not available.}
\centering
%\begin{threeparttable}
\begin{tabular}{|c|c|c|c|c|}
\hline
Composition&$M_w$  [kDa] & $M_h$ [kDa] & $\kappa\,\, (k_B\,T$)&$C_m$ ($\mu{F}/cm^2$)\\\hline\hline
POPC&0.760 &0.448 & 27.8$\,\pm$2.3&$0.72\pm0.04$\\\hline
SOPC&0.787 &0.476 & 30.1$\,\pm$3.1&$0.71\pm0.02$ \\\hline
DOPC&0.786 &0.474 &22.2$\,\pm$2.0&$0.72\pm0.04$ \\\hline
OMPC&0.732 &0.420 &27.1$\,\pm$2.6&$0.71\pm0.03$ \\\hline
DOPC:Chol (1:1)&NA &NA &27.8$\,\pm$4.6 &$0.50\pm0.09$\\\hline
DPPC:Chol (1:1)&NA &NA &121.3$\,\pm$11.0&$0.45\pm0.05$ \\\hline
DOPC:DPPC:Chol (1:1:1)&NA &NA &72.0$\,\pm$8.4&$0.51\pm0.16$ \\\hline
DOPC:DPPC:Chol (1:1:2)&NA &NA &69.2$\,\pm$7.9 &$0.63\pm0.10$\\\hline
PBd$_{13}$-$b$-PEO$_{11}$&1.19 &0.7 & 17.1$\,\pm$1.5&$0.36\pm0.05$ \\\hline
PBd$_{22}$-$b$-PEO$_{14}$&1.80 & 1.35& 31.0$\,\pm$5.1&$0.27\pm0.03$ \\\hline
PBd$_{33}$-$b$-PEO$_{20}$&2.60 & 1.85 & 54.4 $\,\pm$6.4 &$0.23\pm0.04$\\\hline
PBd$_{46}$-$b$-PEO$_{24}$&3.54&2.60& NA&$0.18\pm0.03$\\\hline
PBd$_{54}$-$b$-PEO$_{29}$&4.19&3.10& 154$\pm$ 16.0&$0.18\pm0.04$\\\hline
PBd$_{120}$-$b$-PEO$_{78}$&9.91&6.80& NA&$0.07\pm0.01$\\\hline
\end{tabular}
\label{polymerBR}
\end{table}

\begin{table}[hbt!]
\caption{Membrane viscosities and values of a dye diffusion coefficient (DiC18) for the DOPC:DPPC:Chol ternary system. The values in brackets indicate lipid molar ratios (first column) and the number of measured vesicles (third column). All the experiments were performed at 25.0 $^o$C.  $L_d$ and $L_o$ denote liquid disordered and liquid ordered, respectively. The diffusion coefficient were taken from \cite{SCHERFELD20033758}
}
\centering
%\begin{threeparttable}
\begin{tabular}{|cccc|}
\hline
Multi-component &Phase state&  $\eta_m$ [nPa.s.m]& D [$\mu$m$^2$/s  \cite{SCHERFELD20033758}]\\\hline\hline
DOPC&$L_d$&4.11$\,\pm$2.63 (20)&6.30$\pm$0.13\\
DPPC:Chol (1:1)&$L_o$& 56.4$\,\pm4.63$ (25)&  0.48$\pm$0.06 \\
DOPC:DPPC:Chol (1:1:2)&$L_o$&15.4$\,\pm2.40$ (25)&1.85$\pm$0.13\\
DOPC:DPPC:Chol (1:1:1)&$L_d$&17.7$\,\pm3.06$ (18)& 2.50$\pm$0.20 \\
DOPC:Chol (1:1)& $L_d$& 7.00$\,\pm4.77$ (25)&3.25$\pm$0.25\\
 \hline
 
%PBd$_{22}$-$b$-PEO$_{14}$ &55.7$\,\pm$7.0 (22) CHECK 22 -table 2\\\hline
\end{tabular}
%\end{threeparttable}
 \label{system}
\end{table}

\section{Rheology of block copolymer PBd$_{33}$-$b$-PEO$_{20}$.}

The copolymers PBd$_{33}$-$b$-PEO$_{20}$ was obtained from Polymer Source Inc. (Montreal, Canada). Figure \ref{RHEOLOGYPBDPEO}A and B demonstrate the molecular formula and morphology of the melt at room temperature. The melt appears to be like a pasty wax. The shear rheology response was characterized using Anton MCR rheometer 302 for rheological testing. A parallel plate (PP 25 mm) fixture was used for the measurements for low viscosity measurements. The parallel plate gap width was kept 0.7 to $mm$. 0.5 gm of the sample was loaded on the rheometer. The bulk shear viscosity, $\eta(\dot{\gamma})=\tau_{12}/\dot{\gamma}$, was determined from the measured shear stress, $\tau_{12}$, from imposed shear rates in the range from $\dot{\gamma}=10^{-2}-10^{2}$ s$^{-1}$. As shown in Figure \ref{RHEOLOGYPBDPEO}C, the strain was chosen to be 1 $\%$ to ensure that the sample was measured in linear viscoelastic regime. Figure \ref{RHEOLOGYPBDPEO}D represents the viscosity at different shear rates. The block co polymer demonstrates consistent shear thinning, and total lack of  zero-shear regime. The power law obtained in -0.51. This power law behavior is consistent with other published data in literature \cite{wade2020}.

\begin{figure}[h]
\includegraphics[width=\textwidth]{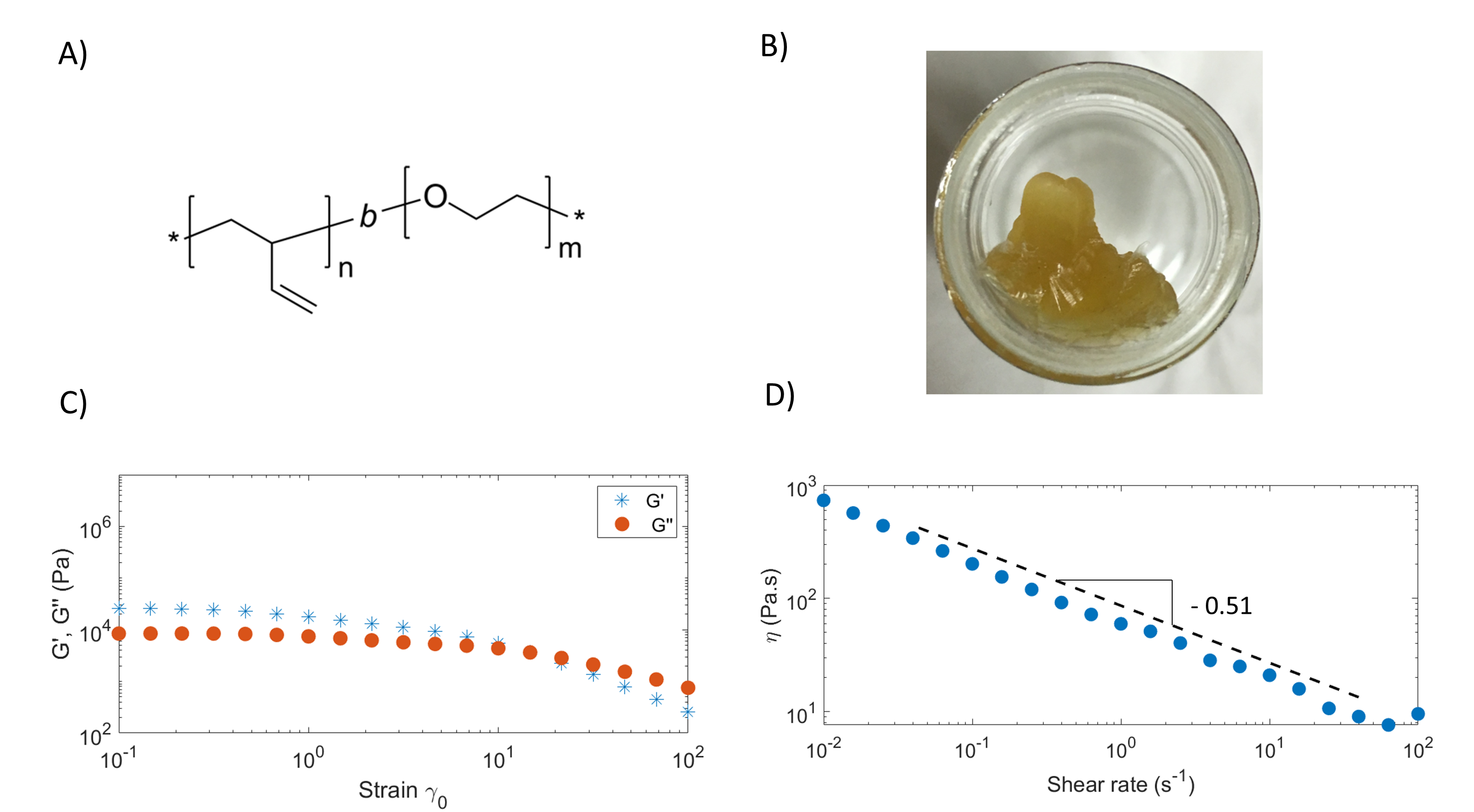}
\centering
\caption{A) Molecular structure of PBd$_{n}$-$b$-PEO$_{m}$ diblock co-polymer. B) The appearance of the polymer melt at room temperature. C) Determination of loss and storage modulus at different strain to determine the linear viscoelastic strain limit. D)  The flow curve experiment with bulk viscosity vs shear rate. }
\centering
\label{RHEOLOGYPBDPEO}
\end{figure}

\end{widetext}

\clearpage

\bibliographystyle{unsrtnat}
%\bibliography{refsVis,refs}

\end{document}